\documentclass[sn-nature, pdflatex]{sn-jnl}

\usepackage{graphicx}%
\usepackage{multirow}%
\usepackage{amsmath,amssymb,amsfonts}%
\usepackage{amsthm}%
\usepackage{mathrsfs}%
\usepackage[title]{appendix}%
\usepackage{xcolor}%
\usepackage{textcomp}%
\usepackage{manyfoot}%
\usepackage{booktabs}%
\usepackage{algorithm}%
\usepackage{algorithmicx}%
\usepackage{algpseudocode}%
\usepackage{listings}%
\usepackage{caption}
\usepackage{subcaption}
\usepackage[superscript]{cite}

\theoremstyle{thmstyleone}%
\theoremstyle{thmstyletwo}%

\theoremstyle{thmstylethree}%

\usepackage{setspace}
\begin{document}
\title[]{Evidence of the quantum-optical nature of high-harmonic generation}

\author*[1]{\fnm{David} \sur{Theidel}}\email{david.theidel@polytechnique.edu}
\author[1]{\fnm{Viviane} \sur{Cotte}}
\author[3, 4]{\fnm{René} \sur{Sondenheimer}}
\author[1]{\fnm{Viktoriia} \sur{Shiriaeva}}
\author[1]{\fnm{Marie} \sur{Froidevaux}}
\author[1]{\fnm{Vladislav} \sur{Severin}}
\author[1]{\fnm{Adam} \sur{Merdji-Larue}}
\author[2]{\fnm{Philip} \sur{Mosel}}
\author[2]{\fnm{Sven} \sur{Fröhlich}}
\author[2]{\fnm{Kim-Alessandro} \sur{Weber}}
\author[2]{\fnm{Uwe} \sur{Morgner}}
\author[2]{\fnm{Milutin} \sur{Kovacev}}
\author[5, 6]{\fnm{Jens} \sur{Biegert}}
\author[1]{\fnm{Hamed} \sur{Merdji}}

\affil[1]{\orgdiv{Laboratoire d’Optique Appliquée}, \orgname{ENSTA ParisTech, CNRS, École polytechnique}, \orgaddress{\street{828 Boulevard des Maréchaux}, \city{Palaiseau}, \postcode{91120}, \country{France}}}
\affil[2]{\orgdiv{Institut für Quantenoptik}, \orgname{Leibniz Universität Hannover}, \orgaddress{\street{Welfengarten 1}, \city{Hanover}, \postcode{30167}, \country{Germany}}}
\affil[3]{\orgname{Fraunhofer Institute for Applied Optics and Precision Engineering IOF}, \orgaddress{\street{Albert-Einstein-Str. 7}, \city{Jena}, \postcode{07745}, \country{Germany}}}
\affil[4]{\orgname{Friedrich-Schiller-University Jena},  \orgname{Institute of Condensed Matter Theory and Optics}, \orgaddress{\street{Max-Wien-Platz 1}, \city{Jena}, \postcode{07743}, \country{Germany}}}
\affil[5]{\orgdiv{ICFO - Institut de Ciencies Fotoniques}, \orgname{The Barcelona Institute of Science and Technology}, \orgaddress{\city{Castelldefels (Barcelona)}, \postcode{08860}, \country{Spain}}}
\affil[6]{\orgdiv{ICREA - Institució Catalana de Recerca i Estudis Avançats}, \orgaddress{\city{Barcelona}, \country{Spain}}}


\abstract{\textbf{High-harmonic generation is a light up-conversion process occurring in a strong laser field, leading to coherent bursts of extreme ultrashort broadband radiation \cite{lewenstein1994theory}. As a new perspective, we propose that ultrafast strong-field electronic or photonic processes such as high-harmonic generation can potentially generate non-classical states of light well before the decoherence of the system occurs \cite{gorlach2020quantum, stammer2022high}. This could address fundamental challenges in quantum technology such as scalability, decoherence or the generation of massively entangled states \cite{lewenstein2022attosecond}. 
Here, we report experimental evidence of the non-classical nature of the harmonic emission in several semiconductors excited by a femtosecond infrared laser. By investigating single- and double beam intensity cross-correlation \cite{loudon1980non}, we measure characteristic, non-classical features in the single photon statistics. We observe two-mode squeezing in the generated harmonic radiation, which depends on the laser intensity that governs the transition from Super-Poissonian to Poissonian photon statistics. The measured violation of the Cauchy-Schwarz inequality realizes a direct test of multipartite entanglement in high-harmonic generation \cite{wasak2014cauchy}. This result is supported by the theory of multimodal detection and the Hamiltonian from which the effective squeezing modes of the harmonics can be derived \cite{gonoskov2022nonclassical, christ2011probing}. 
With this work, we show experimentally that high-harmonic generation is a new quantum bosonic platform that intrinsically produces non-classical states of light with unique features such as multipartite broadband entanglement or multimode squeezing. The source operates at room temperature using standard semiconductors and a standard commercial fiber laser, opening new routes for the quantum industry, such as optical quantum computing, communication and imaging.}}


\maketitle

The study of the quantum properties of light has led to significant advances in basic science and yields a broad impact on modern technology. Since the ground-breaking experiment of Hanbury-Brown-Twiss (HBT) \cite{brown1956correlation} and the theoretical formulation of optical coherence by Glauber \cite{glauber1963quantum}, highly sensitive measurements helped to unravel the non-classical properties of light and provided access to novel methods of sensing \cite{abbott2016observation}, communication \cite{pirandola2020advances}, and imaging \cite{defienne2021polarization, england2019quantum, gilaberte2019perspectives}. Especially single photon emitters, squeezed states of light and entanglement between photons are heavily studied from a fundamental and practical point of view \cite{erhard2020advances}. 
The go-to sources to infer optical non-classical correlations are non-linear interactions of light with the an-harmonic potential in crystals, atoms and artificial atoms \cite{strekalov2019nonlinear}. The platforms are mainly limited to the regime of low photon numbers in the visible spectrum, in which quantum effects are pronounced, limiting the applied perspectives. Recent experimental approaches try to overcome these limitations, e.g., by using high photon number squeezed states \cite{agafonov2010two}, which is crucial for transferring the technology from the lab to real world applications. High-harmonic generation (HHG) in semiconductors using compact fiber laser could lead to these unique properties. In addition to the intrinsic potential of quantum correlation in solid state systems that could be transferred to HHG \cite{alcala2022high}, the on-chip integration of such quantum optical systems is a major factor enabling future photonic quantum technologies. 

In this article, we experimentally investigate the quantum optical properties of HHG in semiconductors as a bright and spectrally broadband (exceeding multi-petahertz) source of non-classical light governed by the coherent strong field dynamics of electrons in the material \cite{gorlach2020quantum, stammer2022high, gonoskov2022nonclassical}. Our studies use modest resources such as industrial-grade semiconductors, a commercial infrared fiber laser, and basic quantum optical detection.
Briefly, HHG is an extremely non-linear process leading to the emission of radiation due to strong field driven recollision and acceleration of carriers in atoms, molecules or solids \cite{mcpherson1987studies, itatani2004tomographic, ghimire2011observation}.
When a strong laser field interacts with a medium, the material’s potential is periodically and anharmonically distorted on a sub-optical cycle timescale. Such ultrafast interaction leads to the emission of attosecond bursts of radiation that can be controlled before strong dephasing and decoherence channels (such as phonons in solids) occur. Depending on the symmetry of the system, one observes odd and even high order harmonics of the fundamental driving field frequency \cite{neufeld2019floquet, ben1993effect}.

Recently, the non-classical signature in the infrared laser beam used to generate high harmonics in gases have been reported as optical Schrödinger cat states \cite{tsatrafyllis2017high,lewenstein2021generation}. However, a direct measurement of the quantum optical nature of the high-harmonic emission has never been observed experimentally. Intense theoretical investigations have started to track the non-classical features of the HHG process \cite{gorlach2020quantum, lewenstein2021generation, gombkotHo2021quantum, stammer2023quantum}. It has been predicted that a non-classical driving field may alter fundamental properties of the harmonic emission, leading for example to a new path for the attosecond control of HHG light through squeezing \cite{even2023photon}. 
Particularly, the presence of multipartite entanglement between an arbitrary number of high-harmonic modes and squeezing effects has been predicted. 
Here, we investigate a reduced system with direct single photon detection.
We focus on the experimental study of the second order intensity correlation function of two adjacent harmonics with quantum optical protocols derived from signal and idler photon statistics studies in the parametric down conversion process (see methods)\cite{fortsch2013versatile}.
The measurement of intensity correlation functions is a powerful method to gain insight on the quantum properties of the source. Especially, the second order correlation function provides direct access to signatures of non-classicality, an important step before realizing quantum state tomography \cite{mandel1995optical}.

\begin{figure}[H]

     \begin{minipage}[t]{\textwidth}
     \raggedright\textbf{a)}
     \includegraphics[width=\linewidth]{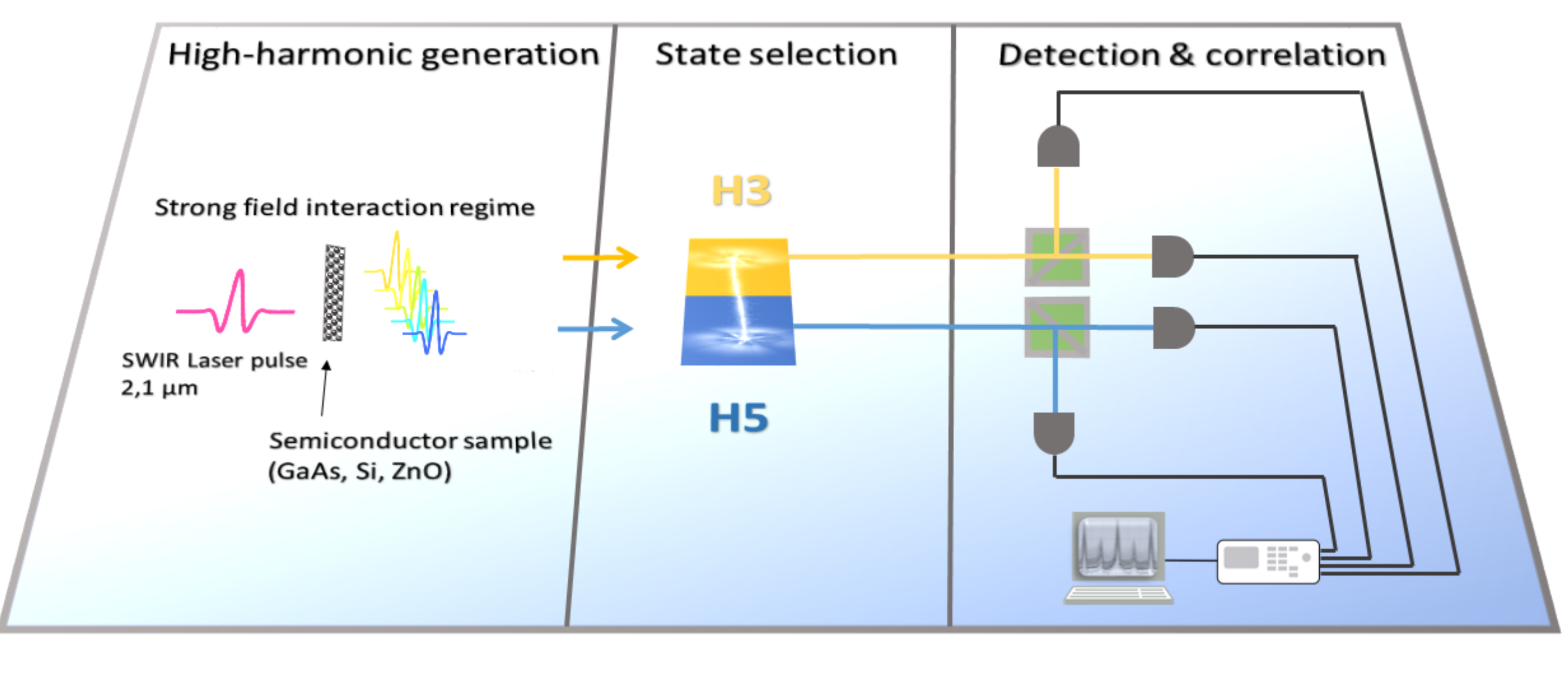}
		\end{minipage}
     \begin{minipage}[t]{0.5\textwidth}
     \raggedright\textbf{b)}

     \includegraphics[width=\linewidth]{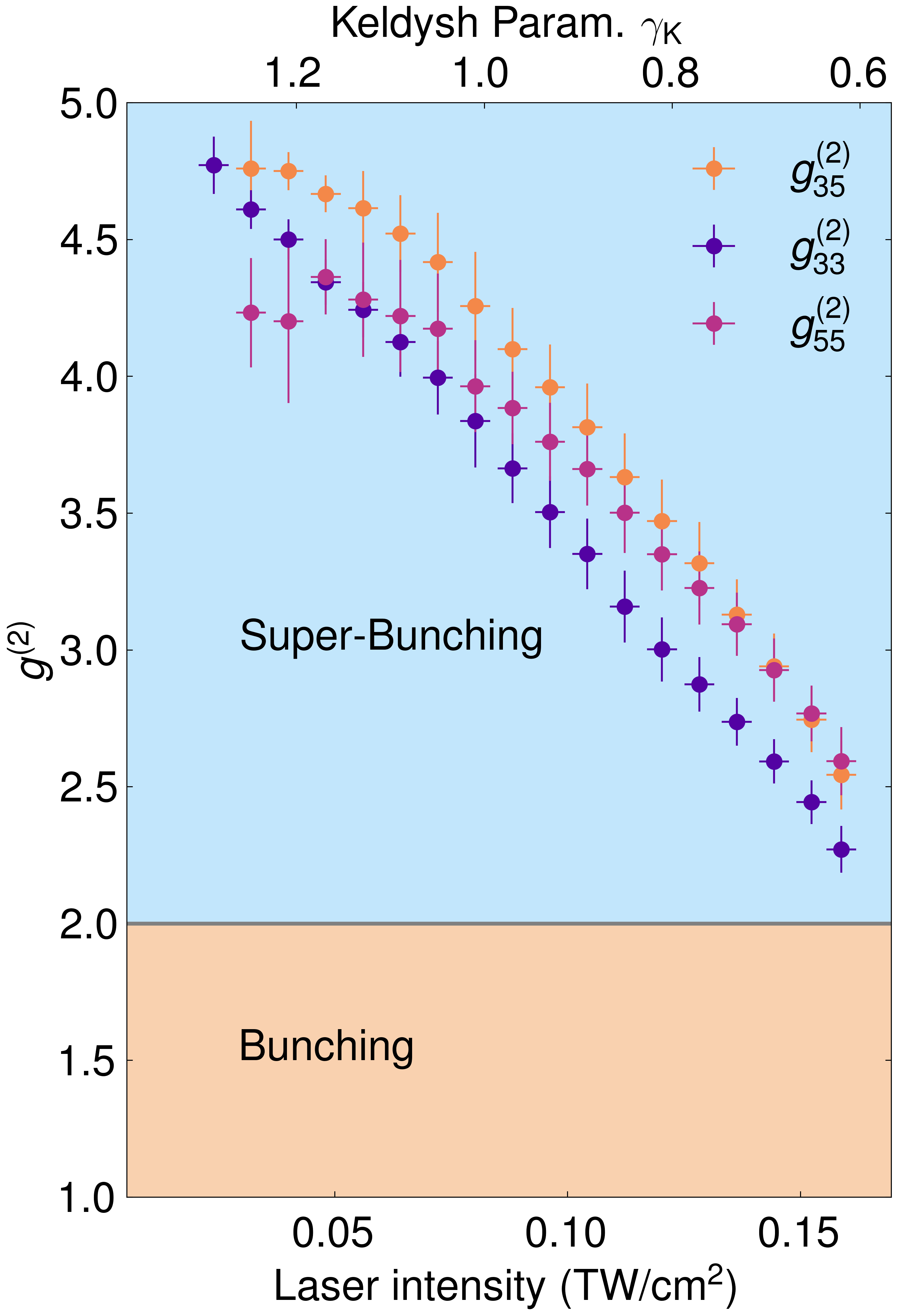}
		\end{minipage}\hfill
	\begin{minipage}[t]{0.5\textwidth}
	\raggedright\textbf{c)}
    \includegraphics[width=\linewidth]{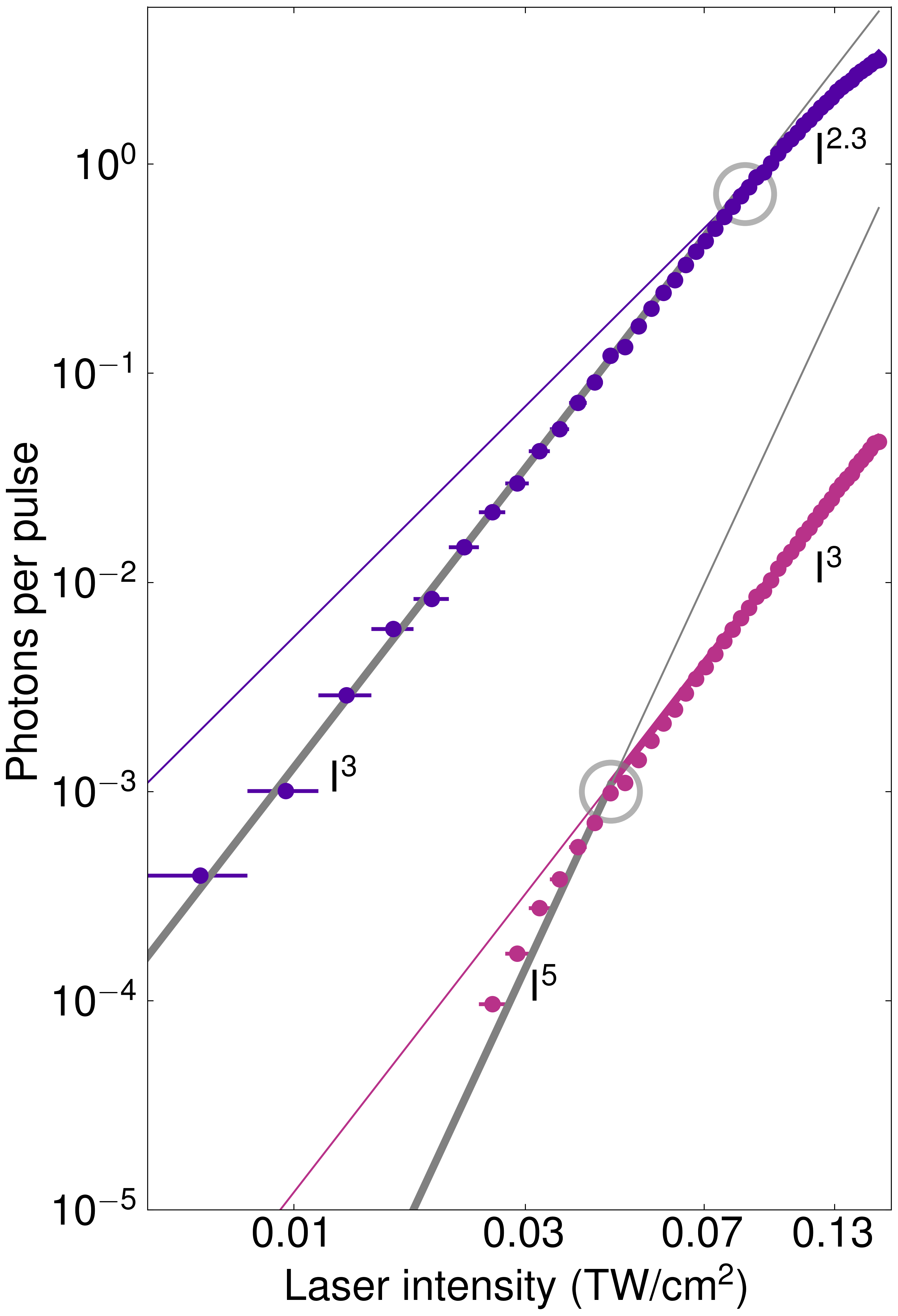}
      \end{minipage}
       \caption{\textbf{Experimental scheme and transition of photon statistics and non-linear intensity dependence of high harmonics for the GaAs sample.} \textbf{a)} The conceptual diagram shows the different experimental stages. The high-harmonic generation process inside the semiconductor sample is driven by ultrafast SWIR laser pulses. Subsequently, the generated bosonic system is reduced to a bipartite system consisting out of two harmonic orders using optical filters for state selection (see supplementary information). The single-photon resolving detection allows to resolve the intensity fluctuations via an intensity correlation measurement. \textbf{b)} Scaling of $g^{(2)}$ with the driving laser intensity, obtained from the correlation of high-harmonic photon counts generated in Gallium Arsenide measured in a single and double beam HBT setup. The blue colored region indicates values of super-bunching with $g^{(2)} > 2$. A decrease of $g^{(2)}$ with increasing laser intensity is recorded for correlations between the same order of high harmonics ($g_{33}^{(2)}$ and $g_{55}^{(2)}$) and for different high-harmonic orders ($g_{35}^{(2)}$). Interestingly, this behaviour is connected to a change in the photon statistics. This behaviour hints on a non-classical state of light generated. The error bars depict the standard error over repetition of the experiment. The scaling Keldysh parameter $\gamma_\mathrm{K}$, given by $\gamma_\mathrm{K} = \omega_\mathrm{L} \sqrt{m^* E_{\mathrm{g}}} / eE$, where $\omega_\mathrm{L}$ is the angular frequency of the driving laser, $m^*$ the reduced electron mass, $E_\mathrm{g}$ the bandgap, $E$ the peak electric field and $e$ the elementary charge \cite{ghimire2014strong}, indicates the interaction regime.  \textbf{c)} Mean number of photons per pulse at the detectors in the HBT setup for H3 (blue) emitted at a central wavelength of 700 nm and at 420 for H5 (purple), calculated by correcting for the SPAD quantum efficiency and losses from optics. With an increase in the driving laser intensity, a non-linear increase of the photon number is observed. The deviation from the perturbative power scaling laws is marked with a circle at the intersection between the perturbative scaling and the best fit, where the exact scaling power law is indicated inside the figure.}
      \label{fig:power_scaling_GaAs}
\end{figure}

Here, we study the photon statistics of HHG photons from the interaction of a short-wavelength infrared (SWIR), ultrashort laser pulse with different semiconductors at room temperature. In detail, we investigate Gallium Arsenide (GaAs) with a  $[100]$ crystal cut, Zinc Oxide (ZnO $[0001]$),and Silicon (Si $[100]$) for the third (H3) and fifth (H5) harmonics. The harmonic measurement, classification and selection is shown in the supplementary information and depicted in Fig. \ref{fig:spectra}. The various band gaps of these semiconductors allow exploring different strong field regimes. 
The employed HBT-like setup sketched in Fig.~\ref{fig:power_scaling_GaAs}a and with more details in Fig.~\ref{fig:setup} measures the second order intensity correlation function given by
\begin{equation}
g^{(2)}_{ij}(\tau) = \frac{\langle a_i^\dagger(\tau) a_j^\dagger(0) a_i(0) a_j(\tau) \rangle}{\langle a_i^\dagger(0) a_i(0) \rangle \langle a_j^\dagger(\tau) a_j(\tau) \rangle}\qquad ,
\label{eq:g2_function}
\end{equation}
where $a^\dagger$ and $a$ denote the creation and annihilation operators of photons in the mode $i$ or $j$ \cite{gombkotHo2021quantum}. If $i = j$ the equation reduces to the standard single-mode normalized second order intensity correlation function (ICF). We will refer also to this, when the indices are omitted. The brackets represent the average over an ensemble. Also, we adopt the notation $g^{(2)}_{ij}(0) = g^{(2)}_{ij}$.

In general, for a single mode bosonic state, the value of $g^{(2)} = 1$ can be identified with Poissonian photon number statistics, as obtained from coherent states. Further, $g^{(2)} > 1$ indicates Super-Poissonian statistics and is connected to bunched arrival of photons at the detector. Non-classical effects like super-bunching result in $g^{(2)} > 2$, while $g^{(2)} < g^{(2)}(\tau)$ is connected to photon antibunching as often obtained from single photon sources.

\begin{figure}
     \centering
     \includegraphics[scale=0.4]{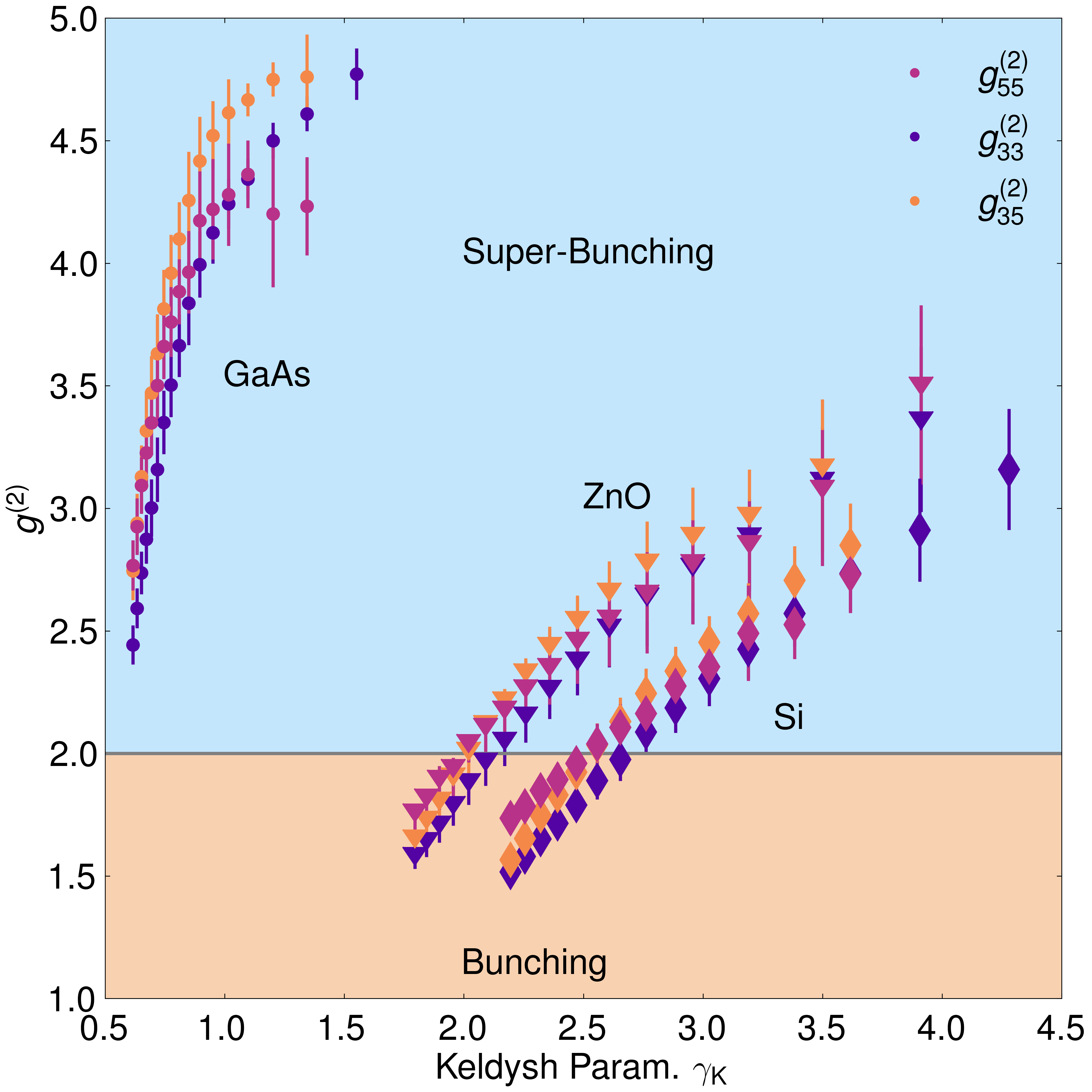}
     \caption{\textbf{Joined results from three different semiconductors.} Experimentally obtained single- and double beam second order correlation values from the third and fifth high-harmonic generated in three different crystals. The data is displayed as a function of the Keldysh parameter. The Keldysh parameter $\gamma_\mathrm{K}$ provides information about the strong-field interaction regime, taking into account the driving laser intensity as well as the reduced electron mass and the bandgap of the material. GaAs shows Super-Poissonian statistics even at the lowest Keldysh parameter. Independent of the generation medium, a transition from Super-Poissonian towards Poissonian photon number statistics is observed, when the Keldysh parameter decreases (i.e. by an increase in the driving laser intensity). The transition in the photon statistics hints on a non-classical state of light generated. The blue colored region indicates values of super-bunching with $g^{(2)} > 2$.}
     \label{fig:g2_keldysh_all}
\end{figure}

We observe a clear dependence of $g^{(2)}$ on the pump power, with a scaling behaviour displayed in Fig.~\ref{fig:power_scaling_GaAs}c in the case of the GaAs semiconductor crystal. Each data point in the graph corresponds to the mean of multiple measurements, and the error bars depict the corresponding standard error.
For low driving laser intensities, a distinct difference to the Poissonian photon number distribution is found in the ICF (Fig. \ref{fig:power_scaling_GaAs}b, Fig. \ref{fig:power_scaling_ZnO}a, Fig. \ref{fig:power_scaling_Si}a). The maximum $g^{(2)}$ values obtained from GaAs indicate super-bunching with $g_{33}^{(2)} = 4.77 \pm 0.10$ and  $g_{55}^{(2)} = 4.36 \pm 0.14$.
The described characteristics are apparent for all investigated semiconductor crystals, while the slope as well as exact maximum and minimum values for $g^{(2)}$ differ (Fig. \ref{fig:g2_keldysh_all}).

Our measurements provide new insight into HHG well beyond the classical Poissonian photon number distribution as usually treated over the last three decades. This is interesting and drives fundamental hints. Indeed, HHG is an inherently coherent process with decoherence channels mainly governed by the electron dynamics of the system, which strongly depend on the laser intensity. 
Instead, we observe for GaAs, ZnO and Si values $g^{(2)} > 1$ for single- and double beam correlations, indicating that the harmonic photons are bunched with Super-Poissonian statistics. Still, the generation process appears to transition towards a Poissonian photon number distribution, as $g^{(2)} = 1$ is approached for higher driving laser intensities. We interpret this observation with the presence of two-mode squeezing in the high-harmonic radiation as described below.

To characterize the harmonic source in terms of non-classical properties, we evaluate the inequality 
\begin{equation}
g^{(2)}_{ii}\cdot g^{(2)}_{jj} < \left[g^{(2)}_{ij}\right]^2
\label{eq:nonclass}
\end{equation}
for a multimode twin-beam state, where the indices $i$ and $j$ refer to the two different harmonic modes. If fulfilled, Eq.~\eqref{eq:nonclass} indicates a non-classical light source, as it corresponds to a violation of the Cauchy-Schwarz inequality (CSI) that is strictly valid for classical correlations \cite{loudon2000quantum, gombkotHo2021quantum}. It is well known that the violation of the CSI by the second-order correlation function indicates the presence of particle entanglement in a many-body system of bosons \cite{wasak2014cauchy, kheruntsyan2012violation}.

\begin{figure}
     \centering
     \includegraphics[scale=0.5]{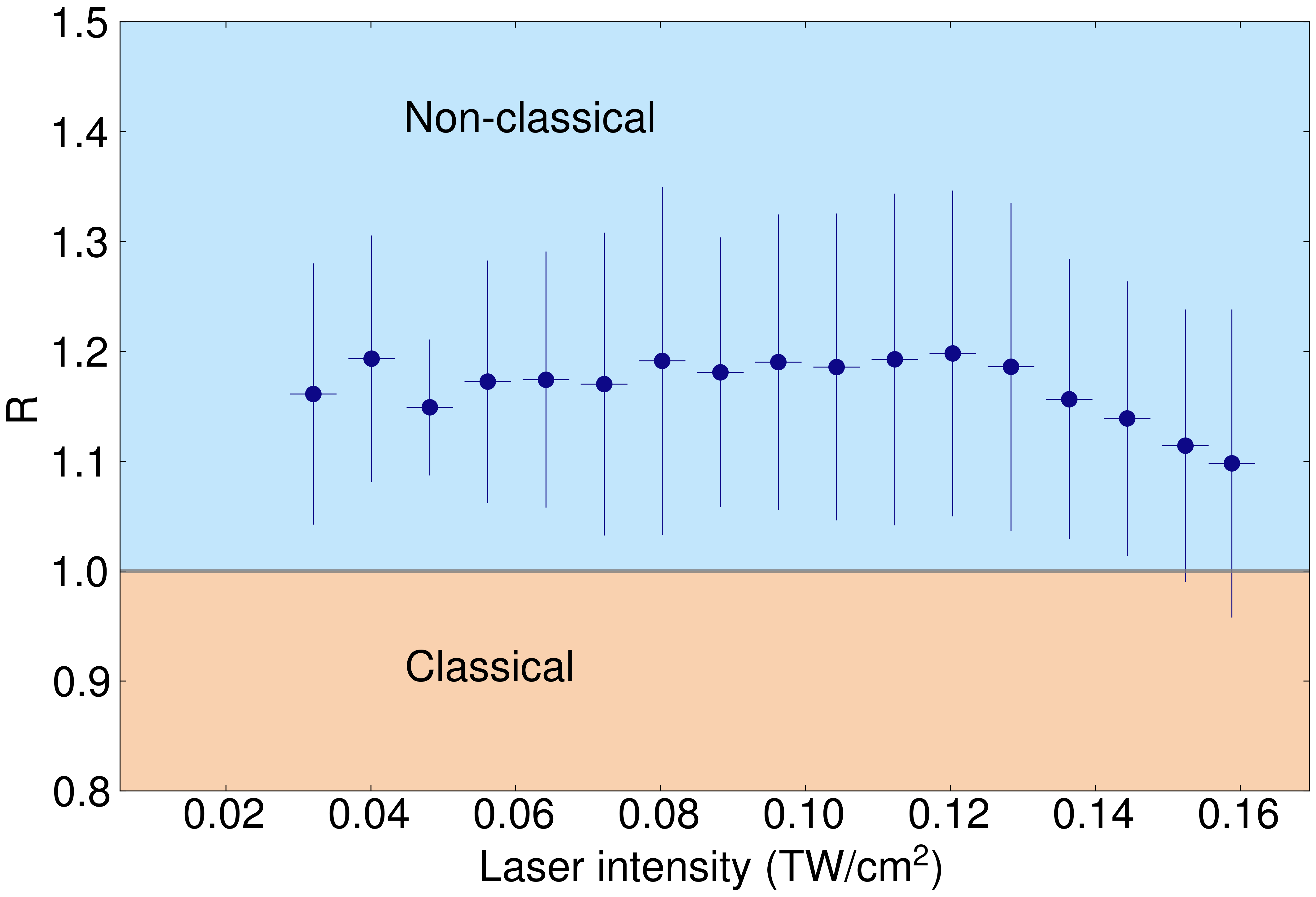}
     \caption{\textbf{Violation of multimode Cauchy-Schwarz inequality.} The graph shows the $R$ parameter, defined as in Eq. \eqref{eq:R_param}, calculated from the experimental data in Fig. \ref{fig:g2_keldysh_all} of the GaAs sample. A value $R > 1$ corresponds to the violation of the classical limit given by the Cauchy-Schwarz inequality. The classical limit is depicted as the gray line. The blue colored region indicates values of non-classicality. The error bars are calculated by error propagation from the original data. The graph shows that the obtained correlation values cross the classical limit of 1, indicating non-classical multimodal correlations. A violation is recorded within one standard deviation.}
     \label{fig:ncl_gaas}
\end{figure}

To this extent, we calculate the $R$ parameter defined as 
\begin{equation}
R := \frac{\left[g^{(2)}_{ij}\right]^2}{g^{(2)}_{ii}\cdot g^{(2)}_{jj}}  
\label{eq:R_param}
\end{equation}
with the experimental data. According to inequality Eq.~\eqref{eq:nonclass}, the bi-partite harmonic emission is non-classical if $R > 1$. 
The results shown in Fig. \ref{fig:ncl_gaas} report the $R$ parameter  calculated from the experimental data in Fig. \ref{fig:power_scaling_GaAs}a.
The graph shows, that the acquired correlation values violate the CSI within one standard deviation. A weaker violation is observed from the other samples (Fig. \ref{fig:ncl_all}).
We interpret that the observed trend of the $R$ parameter with increasing laser intensity is due to the increase of the number of photons in both harmonic modes \cite{loudon2000quantum}.

The presence of squeezing and photon bunching was already predicted theoretically for HHG in atomic targets \cite{gorlach2020quantum, gombkotHo2021quantum, stammer2023quantum} and strongly correlated materials \cite{lange2023electron} and correlated many body systems \cite{pizzi2023light}. Expanding upon these prior studies, our theoretical analysis of HHG in semiconductors shows that harmonic modes can exhibit single-mode and two-mode squeezing.
For our analysis, we adopt the derivation of the quantum properties of intraband harmonics studied in the interaction picture \cite{gonoskov2022nonclassical}.
Here, we perform additional approximations specific to our experimental conditions. We include second order terms with respect to the field operators for the harmonic modes in the interaction Hamiltonian. This is motivated by an effective expansion of the Hamiltonian in terms of the ratio of the operator norm of the field operators of the harmonic fields over the operator norm of the fundamental laser mode. Furthermore, the system is reduced to a bipartite state $|\Psi\rangle_{35}$, consisting of the third and fifth harmonic mode as these two harmonic modes dominate at the reported intensities.

In the interaction regime of the present study, the HHG process is mainly due to intraband excitations, interband transitions are neglected. This is well justified for HHG photons with energies below the materials bandgap, as it is the case for the investigated ZnO and Si samples. For GaAs, photons stemming from interband contributions are expected to additionally contribute to the overall signal.  It is not possible to clearly experimentally differentiate the magnitude of both processes. Since intraband excitations dominate the overall signal \cite{yue2022introduction}, the qualitative predictions made by the model using this approximation remain satisfying. The single particle Hamiltonian describing the interaction is written as 
\begin{equation}
    \hat{H} = n_\mathrm{e} E_\mathrm{c}\left[\hat{\vec{p}} - \frac{e}{c} \hat{\vec{A}}\right] + \omega_3 \hat{n}_3 + \omega_5 \hat{n}_5
\end{equation}
where $e$ the elementary charge, $c$ the speed of light, $n_e$ the time-independent number of electrons in the lowest conduction band, $\hat{\vec{p}}$ the momentum of the electrons, $\hat{n}_j = \hat{a}_j^\dagger \hat{a}_j$ the photon number operator of harmonic mode $j$ with frequency $\omega_j$, $E_c$ the conduction band dispersion and $\hat{\vec{A}} =\vec{A}_{\mathrm{L}} + \hat{\vec{A}}_3 + \hat{\vec{A}}_5$ the vector potential operator of the entire radiation field, where the undepleted driving laser field is treated as a classical field. Treating the canonical momentum of the Bloch electrons as dominated by the driving laser field and switching to the interaction picture yields the
the time-dependent interaction Hamiltonian
\begin{equation}
    \hat{H}_\mathrm{I} = n_\mathrm{e} E_\mathrm{c}\left[-\frac{e}{c} \hat{\vec{A}}(t) \right] \quad .
\end{equation}
We focus on the polarization projection, which allows us to treat the vector potential as a scalar quantity, and approximate the band dispersion as a cosine potential. Substituting the potential operator with the respective field operators, the second order expansion around $-e/c A_\mathrm{L}$ gives
\begin{align}
     \hat{H}_\mathrm{I}(t) = 
     n_\mathrm{e} E_\mathrm{g} \bigg\{ &1 - \cos\left(\frac{\pi \tilde{A}_\mathrm{L}}{K_\mathrm{c}}\right) 
    + \sin\left(\frac{\pi \tilde{A}_\mathrm{L}}{K_\mathrm{c}}\right)  
    \Big[ \tilde{c}_3 \Big(\hat{a}_3(t) + \hat{a}_3^\dagger(t)\Big) + \tilde{c}_5 \Big(\hat{a}_5(t) + \hat{a}_5^\dagger(t)\Big) \Big] \notag\\
    &+  \frac{1}{2} \cos\left(\frac{\pi \tilde{A}_\mathrm{L}}{K_\mathrm{c}}\right) 
    \Big[ \tilde{c}_3^2 \Big(\hat{a}_3(t) + \hat{a}_3^\dagger(t)\Big)^2 +  \tilde{c}_5^2 \Big(\hat{a}_5(t) + \hat{a}_5^\dagger(t)\Big)^2 \notag\\
    &+ 2\tilde{c}_3 \tilde{c}_5 \Big(\hat{a}_3(t) + \hat{a}_3^\dagger(t)\Big) \Big(\hat{a}_5(t) + \hat{a}_5^\dagger(t)\Big)
    \Big] \bigg\} ,
\label{eq:hamiltonian}
\end{align}
 with $\hat{a}_j^\dagger(t)$ and  $\hat{a}_j(t)$ the creation and annihilation operator respectively of high-harmonic mode $j = 3, 5$. Further, $\tilde{c}_j = \pi / K_\mathrm{c} \cdot \sqrt{\pi e^2 / \omega_j V}$ with $K_\mathrm{c}$ an inverse lattice constant, $E_\mathrm{g}$ the conduction band half-width,  $V$ the quantization volume. The classical laser potential is denoted as $\tilde{A}_\mathrm{L} = e/c A_\mathrm{L}$.
An expansion of the two last terms having quadratic dependencies of the field operators in Eq.~\eqref{eq:hamiltonian}, shows the equivalence to an interaction Hamiltonian inducing single or two-mode squeezing in the harmonic modes. 
Based on our theoretical model there are intervals in which the squeezing terms dominate the HHG interaction process while at higher intensities, effects induced by terms linear in the field operators should become stronger. This hints at an optimal operation regime for the source, in which the quantum optical effects are pronounced and could then also occur at higher electric field strength. A detailed analysis of Eq.~\eqref{eq:hamiltonian} is found in the Supplementary Material.

Combining our three main results, the presence of power dependent $g^{(2)}$ values in both single beam and double beam correlations (Fig. \ref{fig:g2_keldysh_all}), as well as violation of the Cauchy-Schwarz inequality (Fig. \ref{fig:ncl_gaas}) and the derived interaction Hamiltonian (Eq. \ref{eq:hamiltonian}), we associate high-harmonic generation with a quantum state that exhibits squeezing.

Photon counting experiments can resolve signatures of squeezing effects, as squeezed states of light characteristically display a change in photon number statistics depending on the squeezing strength and number of squeezing modes as reported in previous work \cite{christ2011probing, walls1983squeezed, cardoso2021superposition}.
The super-bunching for the measured high-harmonics is consistent with related theoretical predictions \cite{gorlach2020quantum}. The observed transition in the photon statistics is a combined result of the dependence of $g^{(2)}_{ij}$ on the squeezing strength \cite{cardoso2021superposition, peng1998introduction} and the multimode, broadband detection over the full harmonic linewidth \cite{christ2011probing}. When the driving laser intensity is increased, the squeezing strength and the number of occupied squeezing modes increases, resulting in $g^{(2)} \rightarrow 1$, as the detection does not discriminate the contribution of the different modes \cite{christ2011probing}. Therefore, a convolution of photon statistics from different modes is effectively measured. 
As the individual modes are not necessarily correlated, this results in a random detection of events with Poissonian statistics. The similarity of the scaling behavior is induced by the dependence of the single- and double beam correlation on the squeezer distribution given by the product of the spectral mode distribution and optical gain. 
In the low gain regime, the value of the single beam correlations depends on the effective number of squeezing modes, whereas the double beam correlation is sensitive to the optical gain of the process \cite{christ2011probing}. These dependencies are the reason for the similar scaling of the measured correlation function between the probed semiconductors.
The apparent reason for this is, that the generation in GaAs has a stronger non-perturbative action illustrated by a lower Keldysh parameter compared to ZnO and Si (see Fig. \ref{fig:ncl_all}). Thus, squeezing effects are less pronounced for ZnO and Si, leading to overall less bunched photons and less correlation between the harmonic modes. This, besides the above mentioned quantum optical effects, contributes to the scaling of $g^{(2)}$ (Fig. \ref{fig:g2_keldysh_all}) and the trend of the $R$ parameter towards the classical limit (Fig. \ref{fig:ncl_gaas}, Fig. \ref{fig:ncl_all}). This finding could be used as a way to engineer photon statistics of the high harmonic emission using various semiconductor, doping or interaction regimes.

\textbf{Conclusion.} We have shown that high-harmonic generation can produce non-classical states of light with unique features such as multipartite broadband entanglement or multimode squeezing. Multimodal non-classical correlations have gained interest in the last decade due to potential applications, however, only few performing and scalable bosonic platforms exist. The HHG source operates at room temperature using standard semiconductors and with commercial femtosecond infrared lasers, opening new routes for the quantum industry such as quantum optical  communication, imaging and computing.
Here, we report experimental evidence of the quantum-optical nature of the high-harmonic emission in various band gap semiconductors. 
On the basis of measuring the second order intensity correlation, we found evidences of a non-classical state of light in the bipartite H3 and H5 harmonics system in GaAs, ZnO and Si. Our experimental results identify two-mode squeezing which is imprinted in the transition from super-Poissonian to Poissonian photon statistics and the violation of the Cauchy-Schwarz inequality. This finding is supported by a theoretical treatment of HHG relative to intraband processes, in accordance with recent theoretical predictions.
We explain the observed change in the HHG photon statistics when increasing the driving laser intensity with a dependence of the photons statistics on the squeezing parameter and the influence of the multimode, broadband detection. In the near future, HHG in the quantum optical regime with a repetition rate in the MHz regime as demonstrated here will provide an ideal platform to produce multipartite entangled photonic systems with high photon number and complex states of light. We have shown that the photon statistics depend on the laser intensity and the semiconductor nature and contains non-classical features, which offers an engineering playground for future applications. Thus, quantum-optical effects cannot be neglected and hints on the possibility to detect non-classical features at higher intensities and higher harmonic orders. With such properties, high-harmonic generation, as a “quantum newcomer”, has interest both from an academic and industrial perspective and the potential to initiate applications in quantum information, communication and sensing.

\backmatter

\bmhead{Supplementary information}

The file contains details on the second order intensity correlation measurement, the classification of the interaction regime, the derivation of the interaction Hamiltonian and Supplementary Figures S1-S8 with legends.

\bmhead{Acknowledgments}
We aknowledge support from Benoît Deveaud at the department of research of Ecole Polytechnique, Philippe Zeitoun and Stéphane Sebban at the Laboratoire d'Optique Appliquée. We acknowledge fruitful discussion with Hugo Defienne from Sorbonne University.


\section*{Declarations}

\subsection*{Funding}
H.M. acknowledges financial support from the European Innovation Council contract EIC open "NanoXCAN" (101047223), ATTOCOM and QUINS contracts from Agence Nationale de la Recherche (ANR).
J.B. acknowledges financial support from the European Research Council for ERC Advanced Grant
“TRANSFORMER” (788218), ERC Proof of Concept Grant “miniX” (840010), FET-OPEN “PETACom” (829153), FET-
OPEN “OPTOlogic” (899794), FET-OPEN “TwistedNano” (101046424), Laserlab-Europe (871124), Marie
Skłodowska-Curie ITN “smart-X” (860553), MINECO for Plan Nacional PID2020–112664 GB-I00, AGAUR for 2017
SGR 1639, MINECO for “Severo Ochoa” (CEX2019-000910-S), Fundació Cellex Barcelona, the CERCA Programme/
Generalitat de Catalunya, and the Alexander von Humboldt Foundation for the Friedrich Wilhelm Bessel Prize.
This work was funded by Deutsche Forschungsgemeinschaft (DFG) under Grant No. KO 3798/11-1. M.K. acknowledges support from the Deutsche Forschungsgemeinschaft (DFG, German Research Foundation) under Germany’s Excellence Strategy within the Cluster of Excellence PhoenixD (EXC 2122, Project ID 390833453) and Quantum Frontiers (EXC-2123, Project ID 390837967).

\subsection*{Contributions}
H.M. and D.T. conceived the study. D.T., V.C., V.Sh. and S.F. performed the experiments. M.K. and K.A.W. provided support on single photon detection. D.T. analyzed the data and provided the plots. R.S. and D.T. conducted the theoretical analysis. D.T. and H.M. wrote the manuscript with input from all authors. All authors discussed the results.

\subsection*{Competing interests}
The authors declare no competing financial interests.

\newpage
\bibliography{sn-bibliography}

\begin{thebibliography}{10}
\expandafter\ifx\csname url\endcsname\relax
  \def\url#1{\burl{#1}}\fi
\expandafter\ifx\csname urlprefix\endcsname\relax\def\urlprefix{URL }\fi
\providecommand{\bibinfo}[2]{#2}
\providecommand{\eprint}[2][]{\url{#2}}
\providecommand{\doi}[1]{\url{https://doi.org/#1}}
\bibcommenthead

\bibitem{lewenstein1994theory}
\bibinfo{author}{Lewenstein, M.}, \bibinfo{author}{Balcou, P.},
  \bibinfo{author}{Ivanov, M.~Y.}, \bibinfo{author}{L’huillier, A.} \&
  \bibinfo{author}{Corkum, P.~B.}
\newblock \bibinfo{title}{Theory of high-harmonic generation by low-frequency
  laser fields}.
\newblock \emph{\bibinfo{journal}{Physical Review A}}
  \textbf{\bibinfo{volume}{49}}, \bibinfo{pages}{2117} (\bibinfo{year}{1994}).

\bibitem{gorlach2020quantum}
\bibinfo{author}{Gorlach, A.}, \bibinfo{author}{Neufeld, O.},
  \bibinfo{author}{Rivera, N.}, \bibinfo{author}{Cohen, O.} \&
  \bibinfo{author}{Kaminer, I.}
\newblock \bibinfo{title}{The quantum-optical nature of high harmonic
  generation}.
\newblock \emph{\bibinfo{journal}{Nature Communications}}
  \textbf{\bibinfo{volume}{11}}, \bibinfo{pages}{4598} (\bibinfo{year}{2020}).

\bibitem{stammer2022high}
\bibinfo{author}{Stammer, P.} \emph{et~al.}
\newblock \bibinfo{title}{High photon number entangled states and coherent
  state superposition from the extreme ultraviolet to the far infrared}.
\newblock \emph{\bibinfo{journal}{Physical Review Letters}}
  \textbf{\bibinfo{volume}{128}}, \bibinfo{pages}{123603}
  (\bibinfo{year}{2022}).

\bibitem{lewenstein2022attosecond}
\bibinfo{author}{Lewenstein, M.} \emph{et~al.}
\newblock \bibinfo{title}{Attosecond physics and quantum information science}.
\newblock \emph{\bibinfo{journal}{arXiv preprint arXiv:2208.14769}}
  (\bibinfo{year}{2022}).

\bibitem{loudon1980non}
\bibinfo{author}{Loudon, R.}
\newblock \bibinfo{title}{Non-classical effects in the statistical properties
  of light}.
\newblock \emph{\bibinfo{journal}{Reports on Progress in Physics}}
  \textbf{\bibinfo{volume}{43}}, \bibinfo{pages}{913} (\bibinfo{year}{1980}).

\bibitem{wasak2014cauchy}
\bibinfo{author}{Wasak, T.}, \bibinfo{author}{Sza{\'n}kowski, P.},
  \bibinfo{author}{Zi{\'n}, P.}, \bibinfo{author}{Trippenbach, M.} \&
  \bibinfo{author}{Chwede{\'n}czuk, J.}
\newblock \bibinfo{title}{Cauchy-schwarz inequality and particle entanglement}.
\newblock \emph{\bibinfo{journal}{Physical Review A}}
  \textbf{\bibinfo{volume}{90}}, \bibinfo{pages}{033616}
  (\bibinfo{year}{2014}).

\bibitem{gonoskov2022nonclassical}
\bibinfo{author}{Gonoskov, I.} \emph{et~al.}
\newblock \bibinfo{title}{Nonclassical light generation and control from
  laser-driven semiconductor intraband excitations}.
\newblock \emph{\bibinfo{journal}{arXiv preprint arXiv:2211.06177}}
  (\bibinfo{year}{2022}).

\bibitem{christ2011probing}
\bibinfo{author}{Christ, A.}, \bibinfo{author}{Laiho, K.},
  \bibinfo{author}{Eckstein, A.}, \bibinfo{author}{Cassemiro, K.~N.} \&
  \bibinfo{author}{Silberhorn, C.}
\newblock \bibinfo{title}{Probing multimode squeezing with correlation
  functions}.
\newblock \emph{\bibinfo{journal}{New Journal of Physics}}
  \textbf{\bibinfo{volume}{13}}, \bibinfo{pages}{033027}
  (\bibinfo{year}{2011}).

\bibitem{brown1956correlation}
\bibinfo{author}{Brown, R.~H.} \& \bibinfo{author}{Twiss, R.~Q.}
\newblock \bibinfo{title}{Correlation between photons in two coherent beams of
  light}.
\newblock \emph{\bibinfo{journal}{Nature}} \textbf{\bibinfo{volume}{177}},
  \bibinfo{pages}{27--29} (\bibinfo{year}{1956}).

\bibitem{glauber1963quantum}
\bibinfo{author}{Glauber, R.~J.}
\newblock \bibinfo{title}{The quantum theory of optical coherence}.
\newblock \emph{\bibinfo{journal}{Physical Review}}
  \textbf{\bibinfo{volume}{130}}, \bibinfo{pages}{2529} (\bibinfo{year}{1963}).

\bibitem{abbott2016observation}
\bibinfo{author}{Abbott, B.~P.} \emph{et~al.}
\newblock \bibinfo{title}{Observation of gravitational waves from a binary
  black hole merger}.
\newblock \emph{\bibinfo{journal}{Physical Review Letters}}
  \textbf{\bibinfo{volume}{116}}, \bibinfo{pages}{061102}
  (\bibinfo{year}{2016}).

\bibitem{pirandola2020advances}
\bibinfo{author}{Pirandola, S.} \emph{et~al.}
\newblock \bibinfo{title}{Advances in quantum cryptography}.
\newblock \emph{\bibinfo{journal}{Advances in optics and photonics}}
  \textbf{\bibinfo{volume}{12}}, \bibinfo{pages}{1012--1236}
  (\bibinfo{year}{2020}).

\bibitem{defienne2021polarization}
\bibinfo{author}{Defienne, H.}, \bibinfo{author}{Ndagano, B.},
  \bibinfo{author}{Lyons, A.} \& \bibinfo{author}{Faccio, D.}
\newblock \bibinfo{title}{Polarization entanglement-enabled quantum
  holography}.
\newblock \emph{\bibinfo{journal}{Nature Physics}}
  \textbf{\bibinfo{volume}{17}}, \bibinfo{pages}{591--597}
  (\bibinfo{year}{2021}).

\bibitem{england2019quantum}
\bibinfo{author}{England, D.~G.}, \bibinfo{author}{Balaji, B.} \&
  \bibinfo{author}{Sussman, B.~J.}
\newblock \bibinfo{title}{Quantum-enhanced standoff detection using correlated
  photon pairs}.
\newblock \emph{\bibinfo{journal}{Physical Review A}}
  \textbf{\bibinfo{volume}{99}}, \bibinfo{pages}{023828}
  (\bibinfo{year}{2019}).

\bibitem{gilaberte2019perspectives}
\bibinfo{author}{Gilaberte~Basset, M.} \emph{et~al.}
\newblock \bibinfo{title}{Perspectives for applications of quantum imaging}.
\newblock \emph{\bibinfo{journal}{Laser \& Photonics Reviews}}
  \textbf{\bibinfo{volume}{13}}, \bibinfo{pages}{1900097}
  (\bibinfo{year}{2019}).

\bibitem{erhard2020advances}
\bibinfo{author}{Erhard, M.}, \bibinfo{author}{Krenn, M.} \&
  \bibinfo{author}{Zeilinger, A.}
\newblock \bibinfo{title}{Advances in high-dimensional quantum entanglement}.
\newblock \emph{\bibinfo{journal}{Nature Reviews Physics}}
  \textbf{\bibinfo{volume}{2}}, \bibinfo{pages}{365--381}
  (\bibinfo{year}{2020}).

\bibitem{strekalov2019nonlinear}
\bibinfo{author}{Strekalov, D.~V.} \& \bibinfo{author}{Leuchs, G.}
\newblock \bibinfo{title}{Nonlinear interactions and non-classical light}.
\newblock \emph{\bibinfo{journal}{Quantum Photonics: Pioneering Advances and
  Emerging Applications}} \bibinfo{pages}{51--101} (\bibinfo{year}{2019}).

\bibitem{agafonov2010two}
\bibinfo{author}{Agafonov, I.~N.}, \bibinfo{author}{Chekhova, M.~V.} \&
  \bibinfo{author}{Leuchs, G.}
\newblock \bibinfo{title}{Two-color bright squeezed vacuum}.
\newblock \emph{\bibinfo{journal}{Physical Review A}}
  \textbf{\bibinfo{volume}{82}}, \bibinfo{pages}{011801}
  (\bibinfo{year}{2010}).

\bibitem{alcala2022high}
\bibinfo{author}{Alcal{\`a}, J.} \emph{et~al.}
\newblock \bibinfo{title}{High-harmonic spectroscopy of quantum phase
  transitions in a high-tc superconductor}.
\newblock \emph{\bibinfo{journal}{Proceedings of the National Academy of
  Sciences}} \textbf{\bibinfo{volume}{119}}, \bibinfo{pages}{e2207766119}
  (\bibinfo{year}{2022}).

\bibitem{mcpherson1987studies}
\bibinfo{author}{McPherson, A.} \emph{et~al.}
\newblock \bibinfo{title}{Studies of multiphoton production of
  vacuum-ultraviolet radiation in the rare gases}.
\newblock \emph{\bibinfo{journal}{JOSA B}} \textbf{\bibinfo{volume}{4}},
  \bibinfo{pages}{595--601} (\bibinfo{year}{1987}).

\bibitem{itatani2004tomographic}
\bibinfo{author}{Itatani, J.} \emph{et~al.}
\newblock \bibinfo{title}{Tomographic imaging of molecular orbitals}.
\newblock \emph{\bibinfo{journal}{Nature}} \textbf{\bibinfo{volume}{432}},
  \bibinfo{pages}{867--871} (\bibinfo{year}{2004}).

\bibitem{ghimire2011observation}
\bibinfo{author}{Ghimire, S.} \emph{et~al.}
\newblock \bibinfo{title}{Observation of high-order harmonic generation in a
  bulk crystal}.
\newblock \emph{\bibinfo{journal}{Nature Physics}}
  \textbf{\bibinfo{volume}{7}}, \bibinfo{pages}{138--141}
  (\bibinfo{year}{2011}).

\bibitem{neufeld2019floquet}
\bibinfo{author}{Neufeld, O.}, \bibinfo{author}{Podolsky, D.} \&
  \bibinfo{author}{Cohen, O.}
\newblock \bibinfo{title}{Floquet group theory and its application to selection
  rules in harmonic generation}.
\newblock \emph{\bibinfo{journal}{Nature Communications}}
  \textbf{\bibinfo{volume}{10}}, \bibinfo{pages}{405} (\bibinfo{year}{2019}).

\bibitem{ben1993effect}
\bibinfo{author}{Ben-Tal, N.}, \bibinfo{author}{Moiseyev, N.} \&
  \bibinfo{author}{Beswick, A.}
\newblock \bibinfo{title}{The effect of hamiltonian symmetry on generation of
  odd and even harmonics}.
\newblock \emph{\bibinfo{journal}{Journal of Physics B: Atomic, Molecular and
  Optical Physics}} \textbf{\bibinfo{volume}{26}}, \bibinfo{pages}{3017}
  (\bibinfo{year}{1993}).

\bibitem{tsatrafyllis2017high}
\bibinfo{author}{Tsatrafyllis, N.}, \bibinfo{author}{Kominis, I.},
  \bibinfo{author}{Gonoskov, I.} \& \bibinfo{author}{Tzallas, P.}
\newblock \bibinfo{title}{High-order harmonics measured by the photon
  statistics of the infrared driving-field exiting the atomic medium}.
\newblock \emph{\bibinfo{journal}{Nature Communications}}
  \textbf{\bibinfo{volume}{8}}, \bibinfo{pages}{15170} (\bibinfo{year}{2017}).

\bibitem{lewenstein2021generation}
\bibinfo{author}{Lewenstein, M.} \emph{et~al.}
\newblock \bibinfo{title}{Generation of optical schr{\"o}dinger cat states in
  intense laser--matter interactions}.
\newblock \emph{\bibinfo{journal}{Nature Physics}}
  \textbf{\bibinfo{volume}{17}}, \bibinfo{pages}{1104--1108}
  (\bibinfo{year}{2021}).

\bibitem{gombkotHo2021quantum}
\bibinfo{author}{Gombk{\"o}t{\H{o}}, {\'A}.}, \bibinfo{author}{F{\"o}ldi, P.}
  \& \bibinfo{author}{Varr{\'o}, S.}
\newblock \bibinfo{title}{Quantum-optical description of photon statistics and
  cross correlations in high-order harmonic generation}.
\newblock \emph{\bibinfo{journal}{Physical Review A}}
  \textbf{\bibinfo{volume}{104}}, \bibinfo{pages}{033703}
  (\bibinfo{year}{2021}).

\bibitem{stammer2023quantum}
\bibinfo{author}{Stammer, P.} \emph{et~al.}
\newblock \bibinfo{title}{Quantum electrodynamics of intense laser-matter
  interactions: A tool for quantum state engineering}.
\newblock \emph{\bibinfo{journal}{PRX Quantum}} \textbf{\bibinfo{volume}{4}},
  \bibinfo{pages}{010201} (\bibinfo{year}{2023}).

\bibitem{even2023photon}
\bibinfo{author}{Even~Tzur, M.} \emph{et~al.}
\newblock \bibinfo{title}{Photon-statistics force in ultrafast electron
  dynamics}.
\newblock \emph{\bibinfo{journal}{Nature Photonics}}
  \textbf{\bibinfo{volume}{17}}, \bibinfo{pages}{501--509}
  (\bibinfo{year}{2023}).

\bibitem{fortsch2013versatile}
\bibinfo{author}{F{\"o}rtsch, M.} \emph{et~al.}
\newblock \bibinfo{title}{A versatile source of single photons for quantum
  information processing}.
\newblock \emph{\bibinfo{journal}{Nature Communications}}
  \textbf{\bibinfo{volume}{4}}, \bibinfo{pages}{1818} (\bibinfo{year}{2013}).

\bibitem{mandel1995optical}
\bibinfo{author}{Mandel, L.} \& \bibinfo{author}{Wolf, E.}
\newblock \emph{\bibinfo{title}{Optical Coherence and Quantum Optics}}
  (\bibinfo{publisher}{Cambridge university press}, \bibinfo{year}{1995}).

\bibitem{ghimire2014strong}
\bibinfo{author}{Ghimire, S.} \emph{et~al.}
\newblock \bibinfo{title}{Strong-field and attosecond physics in solids}.
\newblock \emph{\bibinfo{journal}{Journal of Physics B: Atomic, Molecular and
  Optical Physics}} \textbf{\bibinfo{volume}{47}}, \bibinfo{pages}{204030}
  (\bibinfo{year}{2014}).

\bibitem{loudon2000quantum}
\bibinfo{author}{Loudon, R.}
\newblock \emph{\bibinfo{title}{The quantum theory of light}}
  (\bibinfo{publisher}{OUP Oxford}, \bibinfo{year}{2000}).

\bibitem{kheruntsyan2012violation}
\bibinfo{author}{Kheruntsyan, K.} \emph{et~al.}
\newblock \bibinfo{title}{Violation of the cauchy-schwarz inequality with
  matter waves}.
\newblock \emph{\bibinfo{journal}{Physical Review Letters}}
  \textbf{\bibinfo{volume}{108}}, \bibinfo{pages}{260401}
  (\bibinfo{year}{2012}).

\bibitem{lange2023electron}
\bibinfo{author}{Lange, C.~S.}, \bibinfo{author}{Hansen, T.} \&
  \bibinfo{author}{Madsen, L.~B.}
\newblock \bibinfo{title}{Electron-correlation induced nonclassicallity of
  light from high-harmonic generation}.
\newblock \emph{\bibinfo{journal}{arXiv preprint arXiv:2312.08942}}
  (\bibinfo{year}{2023}).

\bibitem{pizzi2023light}
\bibinfo{author}{Pizzi, A.}, \bibinfo{author}{Gorlach, A.},
  \bibinfo{author}{Rivera, N.}, \bibinfo{author}{Nunnenkamp, A.} \&
  \bibinfo{author}{Kaminer, I.}
\newblock \bibinfo{title}{Light emission from strongly driven many-body
  systems}.
\newblock \emph{\bibinfo{journal}{Nature Physics}}
  \textbf{\bibinfo{volume}{19}}, \bibinfo{pages}{551--561}
  (\bibinfo{year}{2023}).

\bibitem{yue2022introduction}
\bibinfo{author}{Yue, L.} \& \bibinfo{author}{Gaarde, M.~B.}
\newblock \bibinfo{title}{Introduction to theory of high-harmonic generation in
  solids: tutorial}.
\newblock \emph{\bibinfo{journal}{JOSA B}} \textbf{\bibinfo{volume}{39}},
  \bibinfo{pages}{535--555} (\bibinfo{year}{2022}).

\bibitem{walls1983squeezed}
\bibinfo{author}{Walls, D.~F.}
\newblock \bibinfo{title}{Squeezed states of light}.
\newblock \emph{\bibinfo{journal}{Nature}} \textbf{\bibinfo{volume}{306}},
  \bibinfo{pages}{141--146} (\bibinfo{year}{1983}).

\bibitem{cardoso2021superposition}
\bibinfo{author}{Cardoso, F.~R.}, \bibinfo{author}{Rossatto, D.~Z.},
  \bibinfo{author}{Fernandes, G.~P.}, \bibinfo{author}{Higgins, G.} \&
  \bibinfo{author}{Villas-Boas, C.~J.}
\newblock \bibinfo{title}{Superposition of two-mode squeezed states for quantum
  information processing and quantum sensing}.
\newblock \emph{\bibinfo{journal}{Physical Review A}}
  \textbf{\bibinfo{volume}{103}}, \bibinfo{pages}{062405}
  (\bibinfo{year}{2021}).

\bibitem{peng1998introduction}
\bibinfo{author}{Peng, J.-S.} \& \bibinfo{author}{Li, G.-X.}
\newblock \emph{\bibinfo{title}{Introduction to modern quantum optics}}
  (\bibinfo{publisher}{World Scientific}, \bibinfo{year}{1998}).

\bibitem{ivanova2006multiphoton}
\bibinfo{author}{Ivanova, O.~A.}, \bibinfo{author}{Iskhakov, T.~S.},
  \bibinfo{author}{Penin, A.~N.} \& \bibinfo{author}{Chekhova, M.~V.}
\newblock \bibinfo{title}{Multiphoton correlations in parametric
  down-conversion and their measurement in the pulsed regime}.
\newblock \emph{\bibinfo{journal}{Quantum Electronics}}
  \textbf{\bibinfo{volume}{36}}, \bibinfo{pages}{951} (\bibinfo{year}{2006}).

\bibitem{becker2005advanced}
\bibinfo{author}{Becker, W.}
\newblock \emph{\bibinfo{title}{Advanced time-correlated single photon counting
  techniques}} Vol.~\bibinfo{volume}{81} (\bibinfo{publisher}{Springer Science
  \& Business Media}, \bibinfo{year}{2005}).

\bibitem{migdall2013single}
\bibinfo{author}{Migdall, A.}, \bibinfo{author}{Polyakov, S.~V.},
  \bibinfo{author}{Fan, J.} \& \bibinfo{author}{Bienfang, J.~C.}
\newblock \emph{\bibinfo{title}{Single-photon generation and detection: physics
  and applications}}  (\bibinfo{publisher}{Academic Press},
  \bibinfo{year}{2013}).

\bibitem{zielnicki2018joint}
\bibinfo{author}{Zielnicki, K.} \emph{et~al.}
\newblock \bibinfo{title}{Joint spectral characterization of photon-pair
  sources}.
\newblock \emph{\bibinfo{journal}{Journal of Modern Optics}}
  \textbf{\bibinfo{volume}{65}}, \bibinfo{pages}{1141--1160}
  (\bibinfo{year}{2018}).

\end{thebibliography}

\begin{appendices}

\renewcommand{\thefigure}{S\arabic{figure}}
\renewcommand{\theHfigure}{S\arabic{figure}}
\renewcommand{\theequation}{S.\arabic{equation}}
\setcounter{figure}{0}
\newpage

\textbf{Methods.} Our setup, which is sketched in Fig. \ref{fig:setup}, consists out of a $200$ mW commercial all-fiber pulsed laser system with a central wavelength of $2100$ nm, a repetition rate of $18.66$ MHz and a pulse duration of 80 fs. The laser's single mode output beam is linearly polarized and focused with a lens ($\text{F}_1$) of 2.5 cm focal length on the crystalline semiconductor sample, reaching a sub-$\text{TW} / \text{cm}^2$ intensity regime. A half-wave plate followed by a polarizer is employed to adjust the driving laser power. Behind the sample, an IR filter and an aperture is placed to isolate the harmonic radiation from the more divergent driving laser and control the spatial mode of the beam. A broadband polarizer is used, to control the polarization mode of the harmonics and to set the contribution of fluorescence below the detection noise. A lens ($\text{F}_2$) with a focal length of 10 cm focuses the high-harmonic radiation into the detector arms. In the beam path of the harmonic radiation, two dichroic mirrors ($\text{DM}_1$ and $\text{DM}_2$) are inserted to spatially separate H3 and H5 harmonic with central wavelengths of 700 nm and 420 nm, respectively. Behind each dichroic mirror and spectral filters, a HBT-like setup measures the second order intensity correlation function.
The HBT setup consists of a 50/50 beamsplitter with a single photon avalanche diode (SPAD) in each arm. Before each diode, a lens ($F_3$) with a focal length of 5 cm focuses the harmonic radiation onto the sensor area. The SPADs are operated in Geiger and free-running mode with 60\% quantum efficiency for H3 and 40\%  quantum efficiency for H5. Time tags from the detectors are registered with a time-to-digital converter (Time Tagger Ultra, Swabian Instruments) for realizing a start-stop measurement with an electronic timing resolution of 5 ps. All measurements were performed without background subtraction or correction for losses, quantum efficiency or, dark counts.

\section{Supplementary Information}
\begin{figure}[H]
     \begin{minipage}[t]{0.5\textwidth}
     \raggedright\textbf{a)}
     \includegraphics[width=\linewidth]{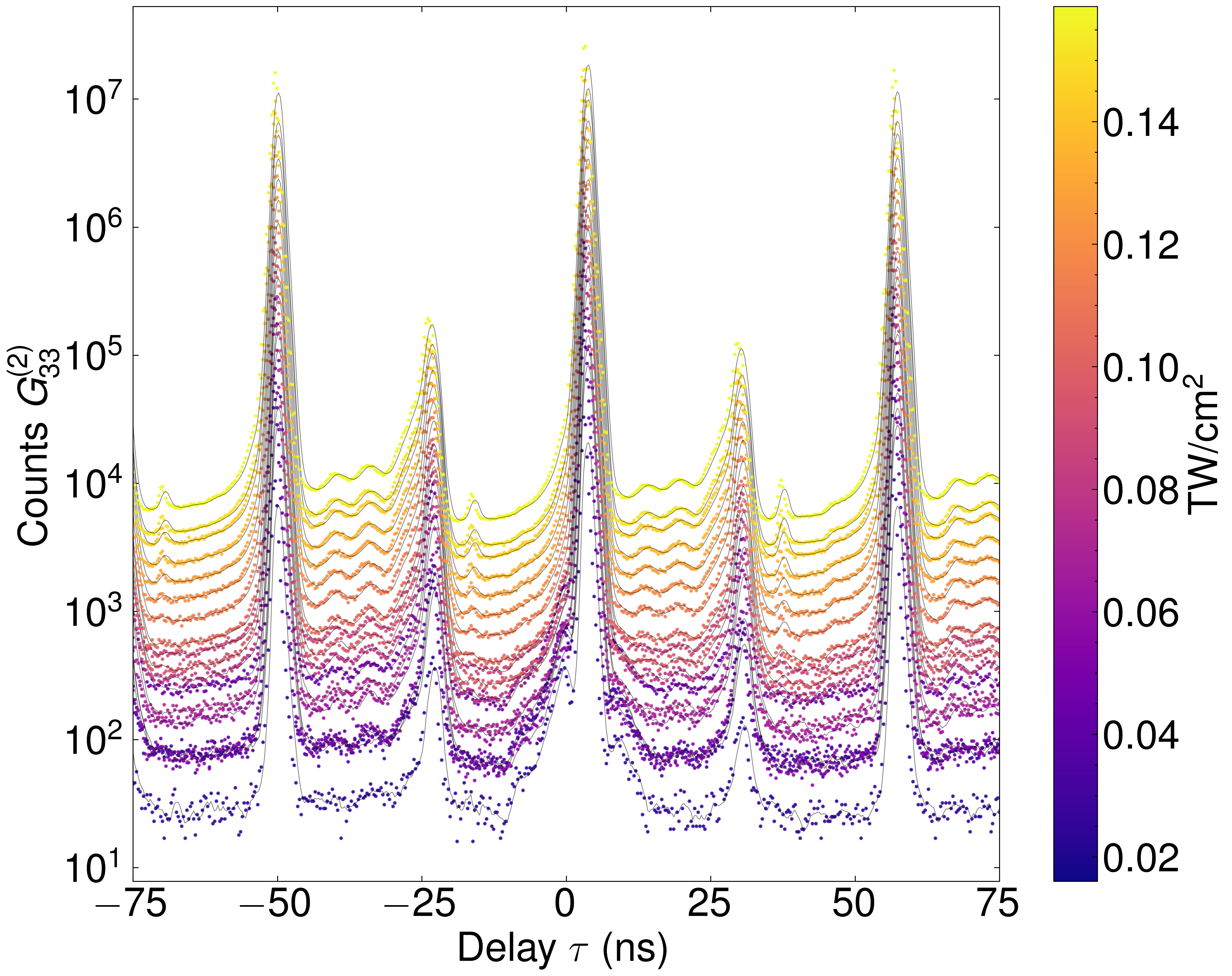}
		\end{minipage}\hfill
	\begin{minipage}[t]{0.5\textwidth}
	\raggedright\textbf{b)}
    \includegraphics[width=\linewidth]{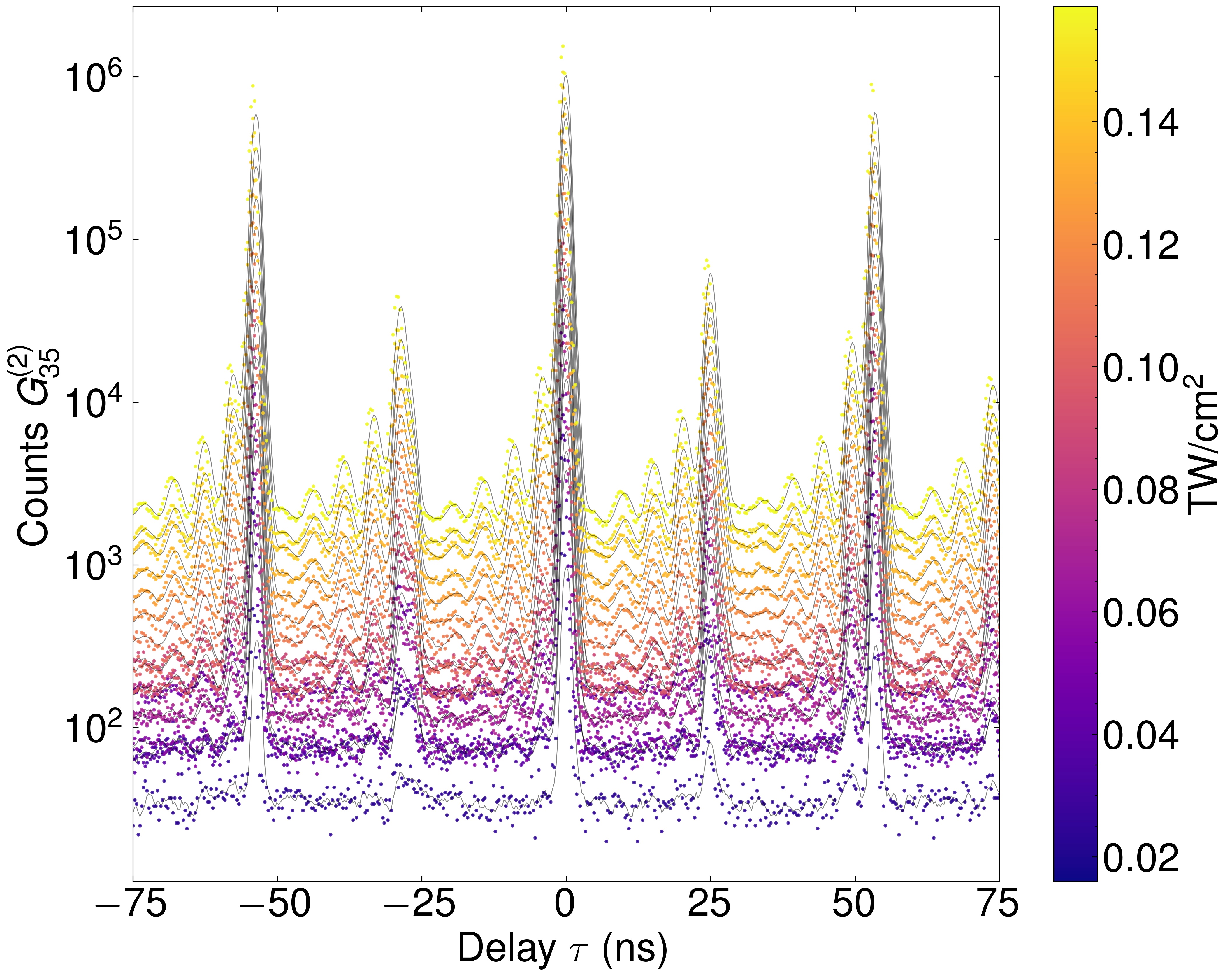}
      \end{minipage}%
       \caption{\textbf{Results of pulsed cross-correlation from GaAs.} \textbf{a)} Single beam cross-correlation measurements on H3 centered at 700 nm generated in GaAs. \textbf{b)} Acquired double beam cross-correlation between H3 and H5 (420 nm). The periodic modulation of the data is due to the pulsed excitation. With increasing driving laser intensity, the overall number of coincidence counts increases. The additional side peaks are the result of electronic noise of the detectors.}
      \label{fig:H3_Counts_Crystals}
 \end{figure}

\textbf{Details on pulsed intensity correlation measurement.}
The data sets (Fig. \ref{fig:H3_Counts_Crystals}) show a periodic modulation due to the pulsed laser excitation \cite{ivanova2006multiphoton}. While the events in the central peak result from photons generated in the same excitation pulse, the satellite peaks for time delays $\tau > 0$ correspond to coincidence counts from photons generated from two subsequent pump pulses. Normalization of the data with respect to the satellite peaks yields the normalized second order correlation function $g^{(2)}(\tau)$ Eq. \eqref{eq:g2_function}.

Figure \ref{fig:H3_Counts_Crystals}a depicts the coincidence counts retrieved on H3 generated in GaAs. The correlation counts in Fig. \ref{fig:H3_Counts_Crystals}b are obtained by correlating the signals from detectors located in the beam paths of H3 and H5.
Noticeably, the number of coincidence counts at zero time delay decreases with respect to the satellite peaks with an increase in pump power. Additionally, we observe an overall increase in the number of coincidence counts with an increase of the driving laser's average power, while the qualitative shape of the function is preserved. 
The background counts in the minima are slightly influenced when doubling the driving laser's power.
The measurements depict a slight asymmetry, shown in Figure \ref{fig:H3_Counts_Crystals}a. The asymmetry is due to photons absorbed in the neutral region of the SPAD sensor. This leads to a delayed registration of photon events, as the electron avalanche in the sensor is triggered by an electron diffusing from the neutral to the active region of the sensor. Also, additional side peaks in Figure \ref{fig:H3_Counts_Crystals}b between the central peak and the satellite peaks originate from after-pulsing of the detectors.  The after-pulsing is also responsible for the slight increase of the background with increasing signal \cite{becker2005advanced}.
The normalization of the raw data displayed in Fig. \ref{fig:H3_Counts_Crystals} is performed with respect to the satellite peaks. As these are uncorrelated, we obtain $g^{(2)}(\tau) = 1$ \cite{ivanova2006multiphoton}. The normalization is done by fitting a Gaussian function to the satellite peak. The peak value of it serves as the reference for normalization. Additionally, data with a coefficient of determination less than $0.95$ is not evaluated. This is especially relevant for the data points obtained at low driving laser intensities including the fifth harmonic.

Using a pulsed source and binary detectors, the measurement of count correlations is in general not equal to the normalized second order ICF. The intensity of the harmonics is low and the detection efficiency of the SPAD is limited, we record predominantly single photon events and thus get a good approximate of the value of $g^{(2)}$ \cite{migdall2013single}.

\begin{figure}[H]
     \begin{minipage}[t]{0.5\textwidth}
     \raggedright\textbf{a)}
     \includegraphics[width=\linewidth]{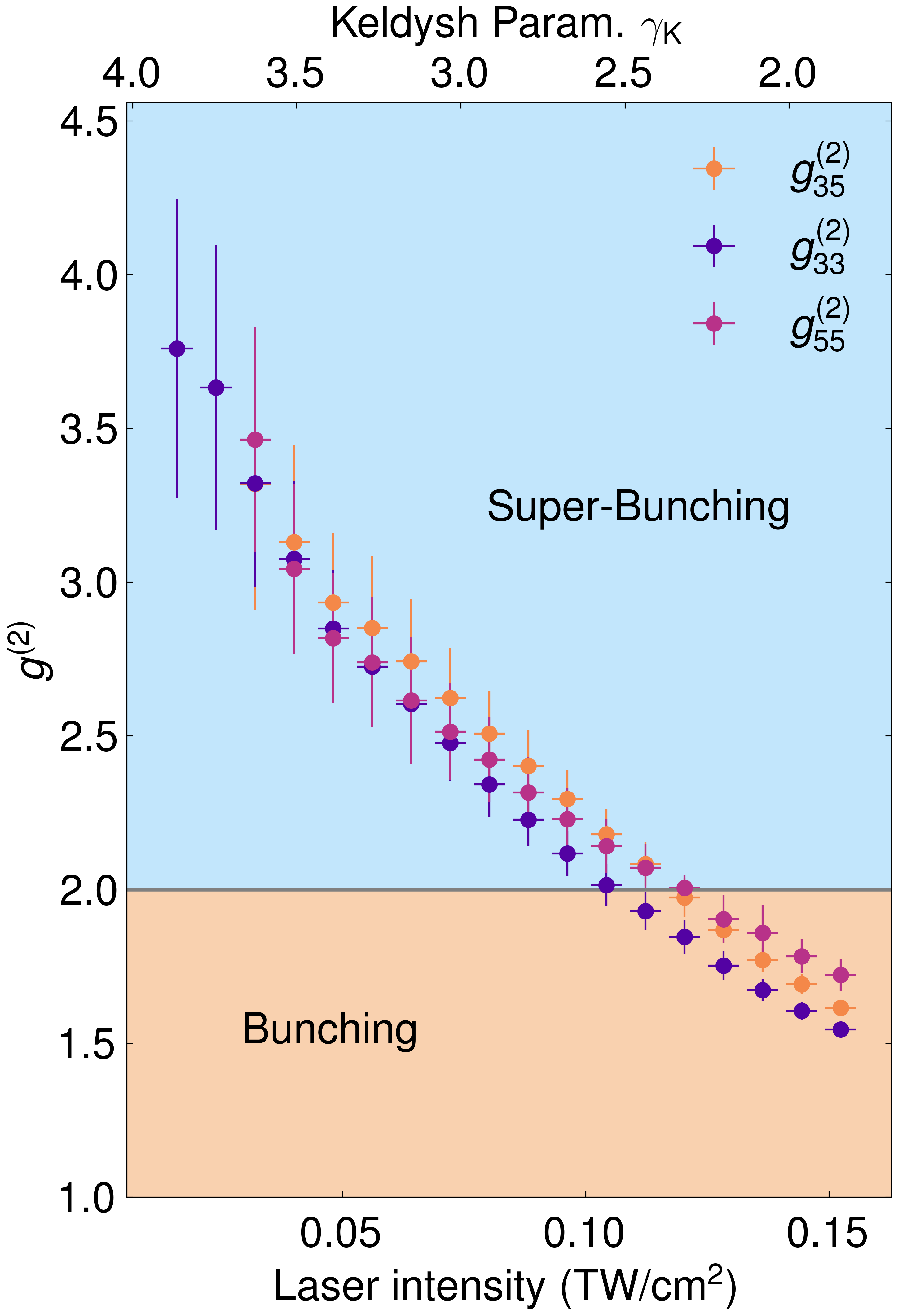}
		\end{minipage}\hfill
	\begin{minipage}[t]{0.5\textwidth}
	\raggedright\textbf{b)}
    \includegraphics[width=\linewidth]{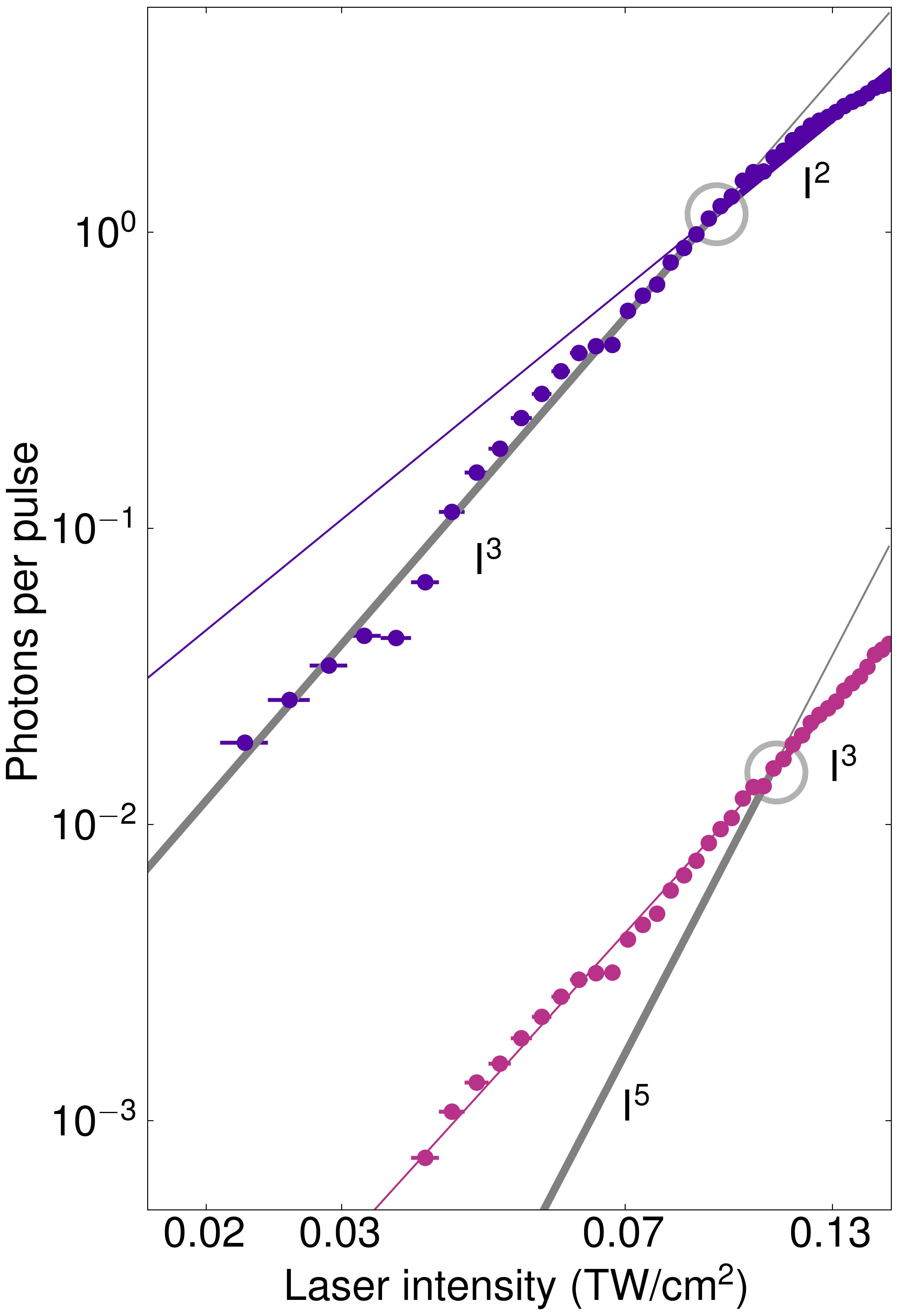}
      \end{minipage}%
       \caption{\textbf{Transition of photon statistics and non-linear intensity dependence of perturbative harmonics in ZnO.} \textbf{a)} Scaling of $g^{(2)}$ obtained from the correlation of high-harmonic photon counts generated in ZnO, measured in a single and double beam HBT setup. A decrease of $g^{(2)}$ with increasing laser intensity is recorded for correlations between the same order of harmonics ($g_{33}^{(2)}$ and $g_{55}^{(2)}$) and for different harmonic orders ($g_{35}^{(2)}$). This change is connected with a change from Super-Poissonian to Poissonian number statistics. This result is interesting as the generation of high harmonics is assumed to be a coherent process.  The error bars depict the standard error over repetition of the experiment. The blue colored region indicates values of super-bunching with $g^{(2)} > 2$. \textbf{b)} Mean number of harmonic photons per pulse is shown over the laser intensity. For H3 (blue dots) and H5 (purple) dots the intensity cross-over is visible, showing the transition from the perturbative to the non-perturbative regime.}
      \label{fig:power_scaling_ZnO}
\end{figure}

\begin{figure}[H]
     \begin{minipage}[t]{0.5\textwidth}
     \raggedright\textbf{a)}
     \includegraphics[width=\linewidth]{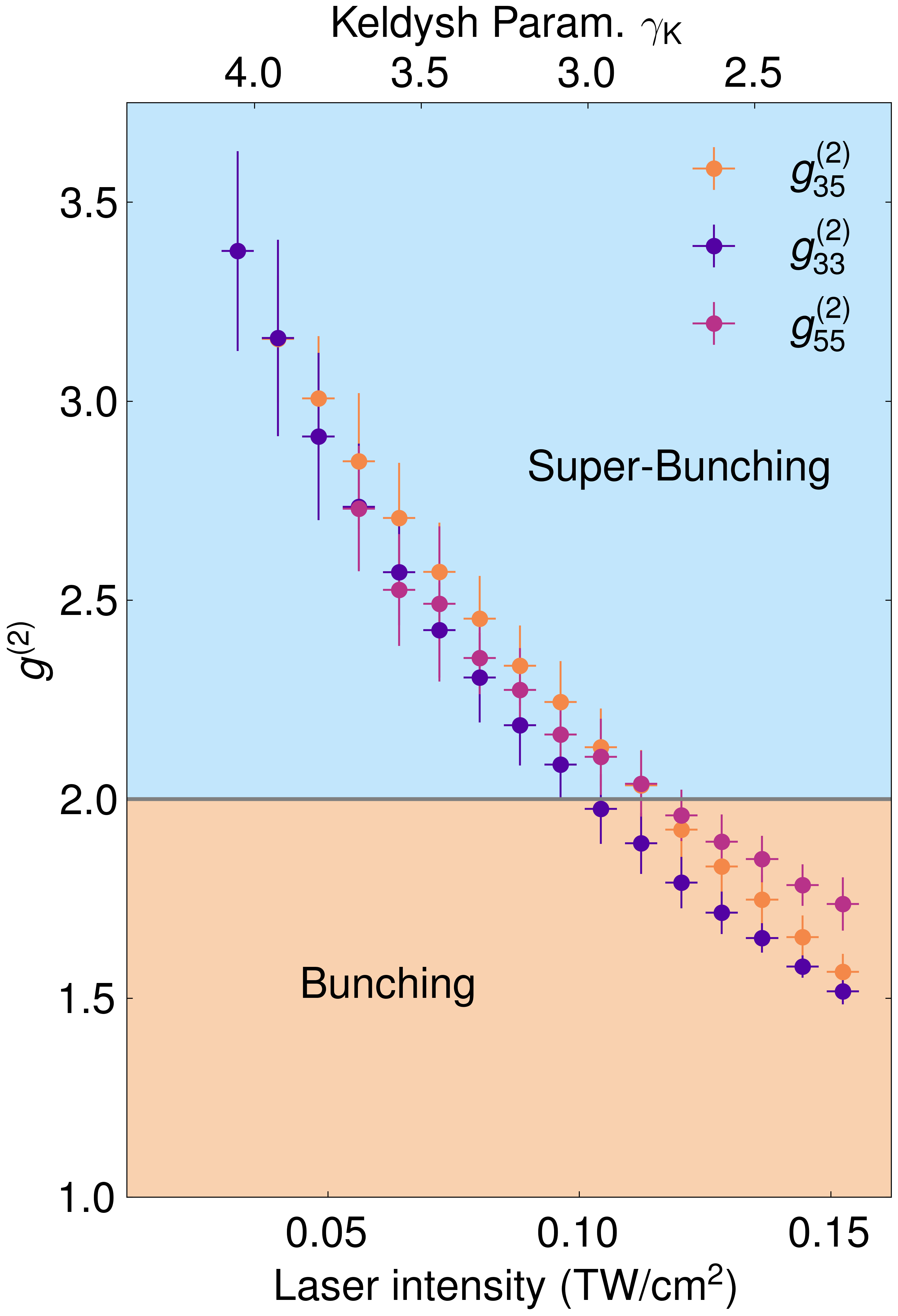}
		\end{minipage}\hfill
	\begin{minipage}[t]{0.5\textwidth}
	\raggedright\textbf{b)}
    \includegraphics[width=\linewidth]{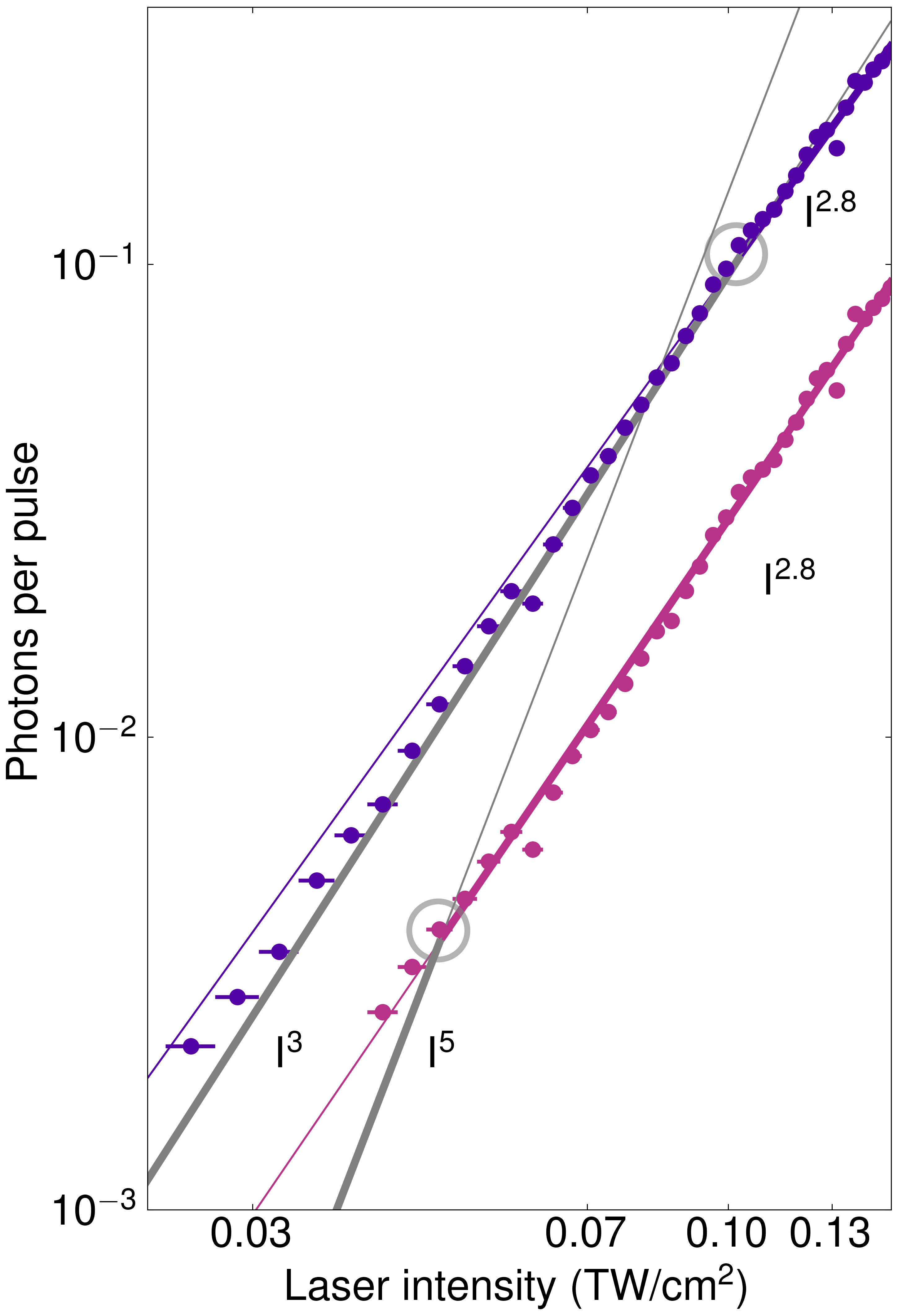}
      \end{minipage}%
       \caption{\textbf{Transition of photon statistics and non-linear intensity dependence of perturbative harmonics in Si.} \textbf{a)} Similar to Fig. \ref{fig:power_scaling_GaAs} the scaling of $g^{(2)}$ obtained from the correlation of high-harmonic photon generated in Si is shown. The same trends for $g^{(2)}$ with increasing laser intensity are observed for correlations between the same order of higher harmonics ($g_{33}^{(2)}$ and $g_{55}^{(2)}$) and for different high-harmonic orders ($g_{35}^{(2)}$). The blue colored region indicates values of super-bunching with $g^{(2)} > 2$. \textbf{b)} Mean number of harmonic photons per pulse is shown over the laser intensity. For H3 (blue dots) and H5 (purple) dots the intensity cross-over is visible, showing the transition from the perturbative to the non-perturbative regime.}
      \label{fig:power_scaling_Si}
\end{figure}

\begin{figure}[H]
     \centering

     \begin{minipage}[t]{0.5\textwidth}
     \raggedright\textbf{a)}
     \includegraphics[width=\linewidth]{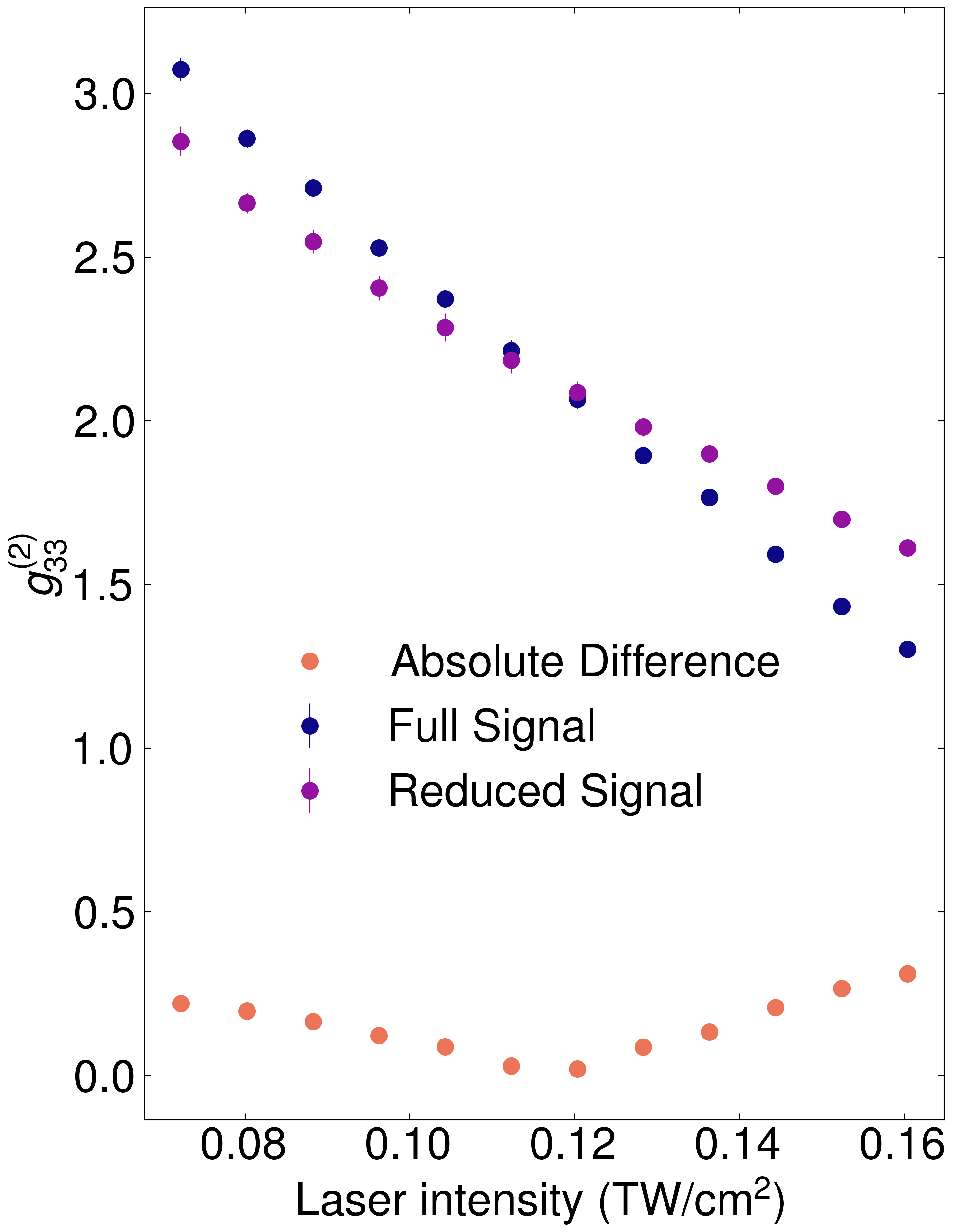}
		\end{minipage}\hfill
	\begin{minipage}[t]{0.5\textwidth}
	\raggedright\textbf{b)}
    \includegraphics[width=\linewidth]{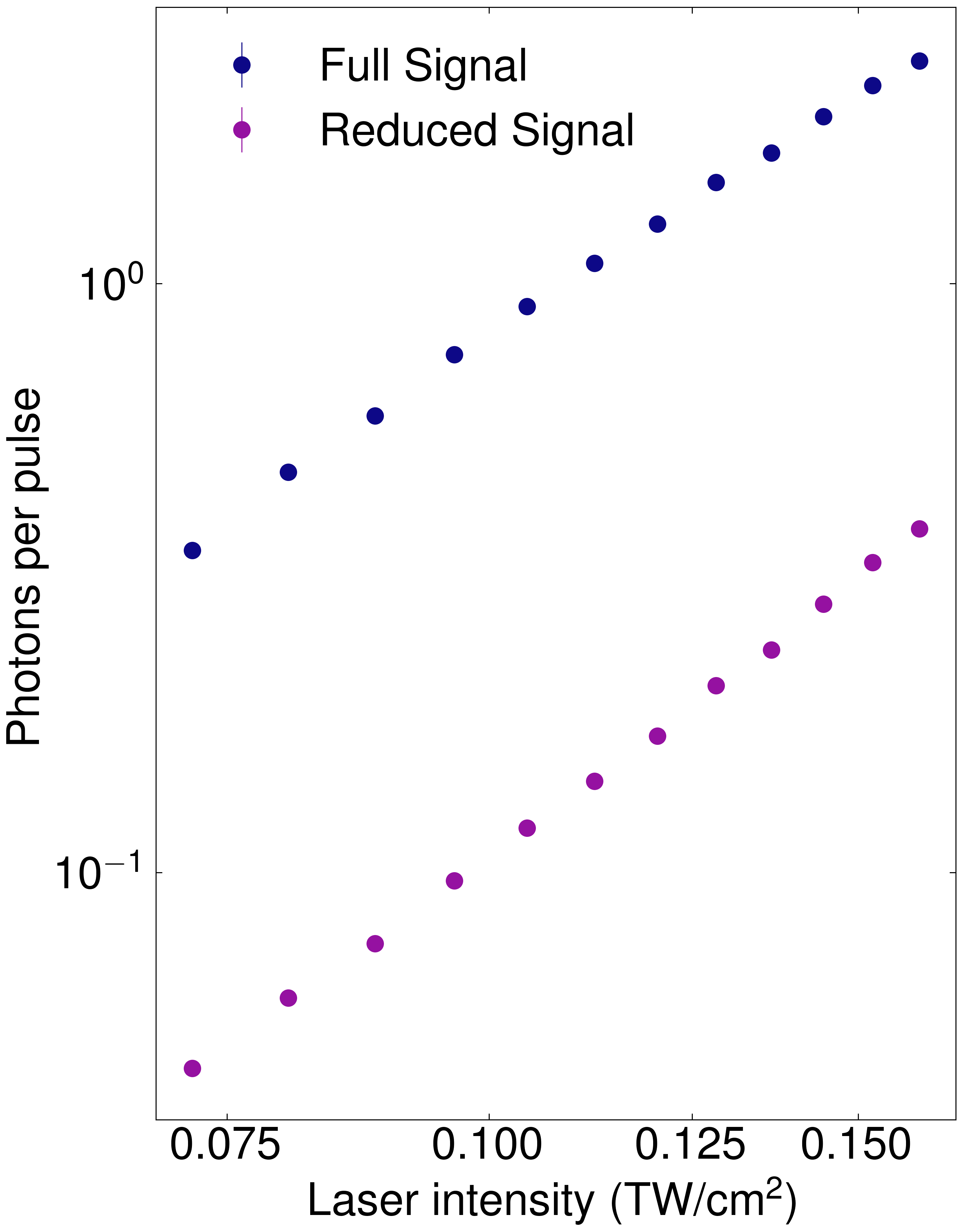}
      \end{minipage}%
     \centering
	
     \caption{\textbf{Loss-tolerant test of photon statistics measurement.} \textbf{a)} Measured value of the second order correlation $g^{(2)}$ for the third high-harmonic generated in GaAs. For the measurement with the reduced harmonic signal, a second polarizer is placed after the sample to introduce additional losses. This results in a reduced photon per pulse yield by one order of magnitude a shown in \textbf{b)}, where the harmonic yield is depicted over the driving laser intensity. As the $g^{(2)}$ measurement is tolerant to losses, the obtained  $g^{(2)}$ values should be reproduced. The test is performed on the signal of H3 at high intensities, where multi-event effects could possibly distort the measurement due to the intense photon yield. We observe that the  $g^{(2)}$ values as well as the trend in the photon statistics is preserved. The absolute difference between the two measurements is below 20 $\%$.}
\label{fig:loss_test}
\end{figure}

We estimate the mean number of photons with
\begin{equation}
    \mu = \frac{N_{\mathrm{H}}}{R_{\mathrm{p}} \cdot \eta}
\end{equation}
where $N_{\mathrm{H}}$ is the count rate of the high-harmonic photons after correction for dark counts, $R_{\mathrm{p}}$ the repetition rate of the laser and $\eta$ the quantum efficiency of the detector. 
For the most intense H3 signal using the measured count rate from Si (Fig. \ref{fig:power_scaling_Si}b) and a quantum efficiency of 0.6, we obtain ${\mu = 0.91}  < 1$. For the ZnO and the GaAs sample, ${\mu = 0.75}$ and, $\mu = 0.49$ respectively. With the decrease of the driving laser intensity, the countrate drops quickly and $\mu \ll 1$ for the great majority of data points for all samples (Fig. \ref{fig:power_scaling_ZnO}b, Fig. \ref{fig:power_scaling_GaAs}b, Fig.  \ref{fig:power_scaling_Si}b). Additionally, the ratio of the coincidence countrate $N_\mathrm{C}$ to the photon countrate $N_\mathrm{H}$ evaluates to $N_\mathrm{C} / N_\mathrm{H} \leq 2 \cdot 10^{-2}$ for all materials at all driving laser intensities. Thus, $g^{(2)}$ can be measured well using Geiger mode, single photon detectors. 
To further verify our measurement, we perform a loss-tolerant test. As the second order correlation function $g^{(2)}$ is tolerant to losses, an additional introduced optical loss should not distort the measurement, unless there is a significant contribution of multi-event effects. An attenuation of the intense H3 signal in GaAs by one order of magnitude preserves the obtained values and the observed trend in the photon statistics (Fig. \ref{fig:loss_test}). Thus, as indicated by the two parameters above and the loss-tolerance of the measurement, multi-event effects do not influence our data. Additional correction for optical losses yields the maximum generation rate in Si at $1.7\cdot 10^{7}$ photons per second for H3 and $1\cdot 10^{6}$ photons per second for H5 during the correlation experiment.

The maximum value for the single beam correlation on H3 in GaAs, $g_{33}^{(2)} = 4.77 \pm 0.10$,  indicates super-bunching. 
Similarly, super-bunching is observed for Si with $g_{33}^{(2)} = 3.38 \pm 0.25$ and for ZnO with $g_{33}^{(2)} = 3.84 \pm 0.48$. 
The single beam correlation on H5 reaches its maximum value $g_{55}^{(2)} = 4.36 \pm 0.14$ in GaAs only for a higher driving laser intensity, and falls off with decreasing laser intensity. 
For the Si sample $g_{55}^{(2)} = 2.5 \pm 0.15$ and the ZnO sample $g_{55}^{(2)} = 3.03 \pm 0.28$. The respective maxima for the double beam correlation are $g_{35}^{(2)} = 4.75 \pm 0.07$ in GaAs {(Si  ${g_{35}^{(2)} = 2.69 \pm 0.13}$, ZnO $g_{35}^{(2)} = 3.12 \pm 0.29$)}.
The scaling of the double beam correlation $g_{35}^{(2)}$ is similar to the single beam correlation $g_{33}^{(2)}$.

We note, that the strong field regime of HHG in semiconductors is usually accompanied by the generation of fluorescence light. The emission bands of the fluorescence light are thereby determined by the bandgap of the semiconductor and the excitation wavelength. In some cases, these are close to the harmonic spectral modes and may influence the photon statistics measurement. In our work, by locating a narrow band light polarizer behind the sample, we discard most of the isotropic unpolarized fluorescence. Finally, the fluorescence signal contributes less than 5$\%$ to our signal. This value is within the dark counts of the detectors, and it was confirmed by rotating the polarizer orthogonal to the linearly polarized high harmonics.

\textbf{Classification and selection of high-harmonic state generation.}

From the registered count rates, we estimate the number of photons per pulse by taking into account the quantum efficiency of the SPAD and the losses introduced by the optics for the respective detector arm.
The number of photons per pulse increases non-linearly with the driving laser intensity. Depending on the generation mechanism of the high-harmonic states, the relationship between the harmonic intensity and the driving laser intensity follows a different scaling law. For perturbative harmonics the exponent of the power law is equal to the harmonic order, whereas the scaling law for non-perturbative harmonics deviates from this. In our measurements, the scaling of the harmonic photon yield for H5 shows clear non-perturbative characteristics for all crystals. For H3, a smaller deviation from the perturbative scaling law is apparent (Fig. \ref{fig:power_scaling_GaAs}b, Fig. \ref{fig:power_scaling_ZnO}b and Fig. \ref{fig:power_scaling_Si}b).

\begin{figure}[H]
     \centering
     \begin{minipage}[t]{0.7\textwidth}
        \begin{subfigure}{\textwidth}
        \subcaption{}
       \includegraphics[width=\linewidth]{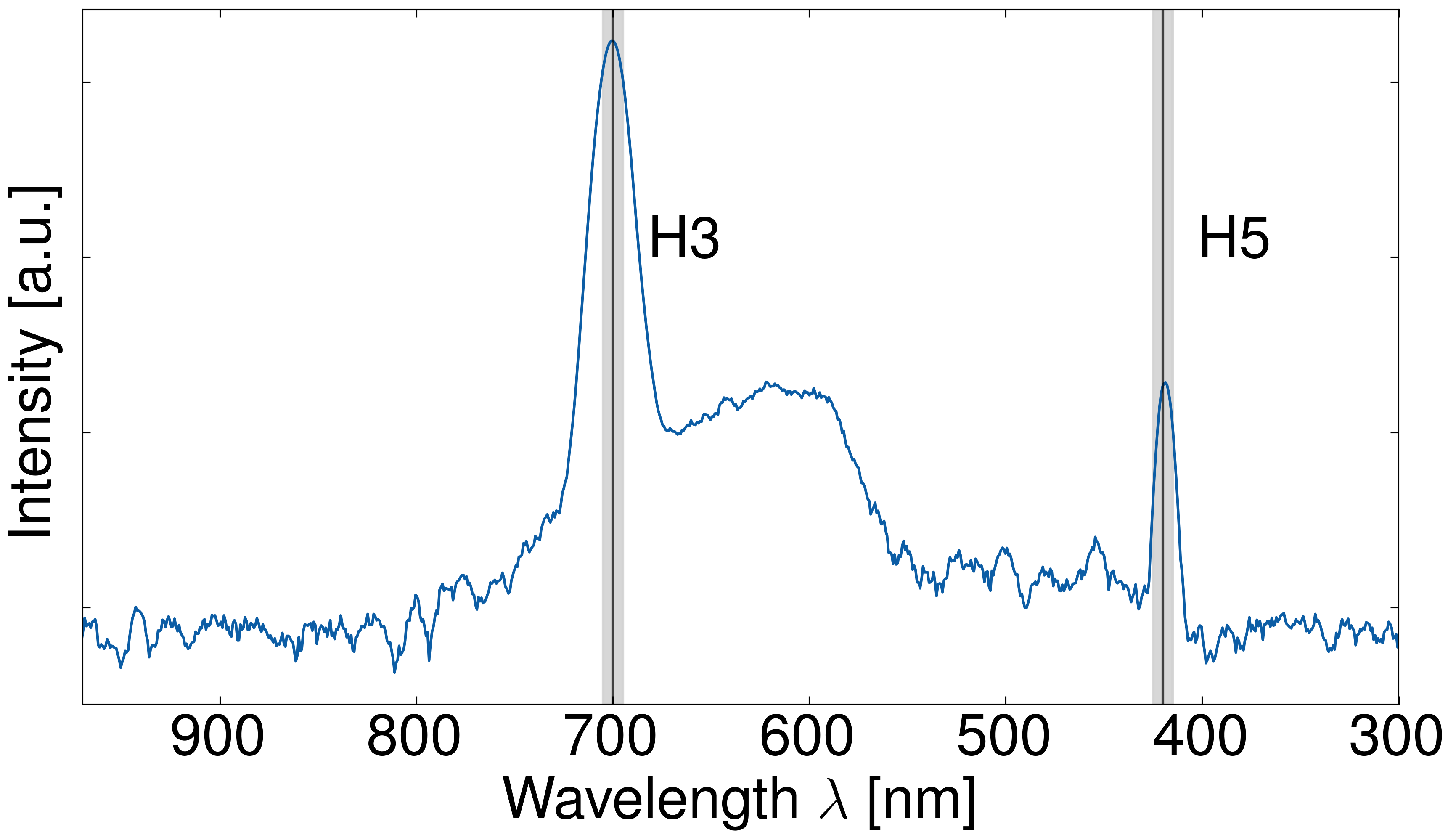}
       \end{subfigure}\par
       \begin{subfigure}{\textwidth}
       \subcaption{}
            \includegraphics[width=\linewidth]{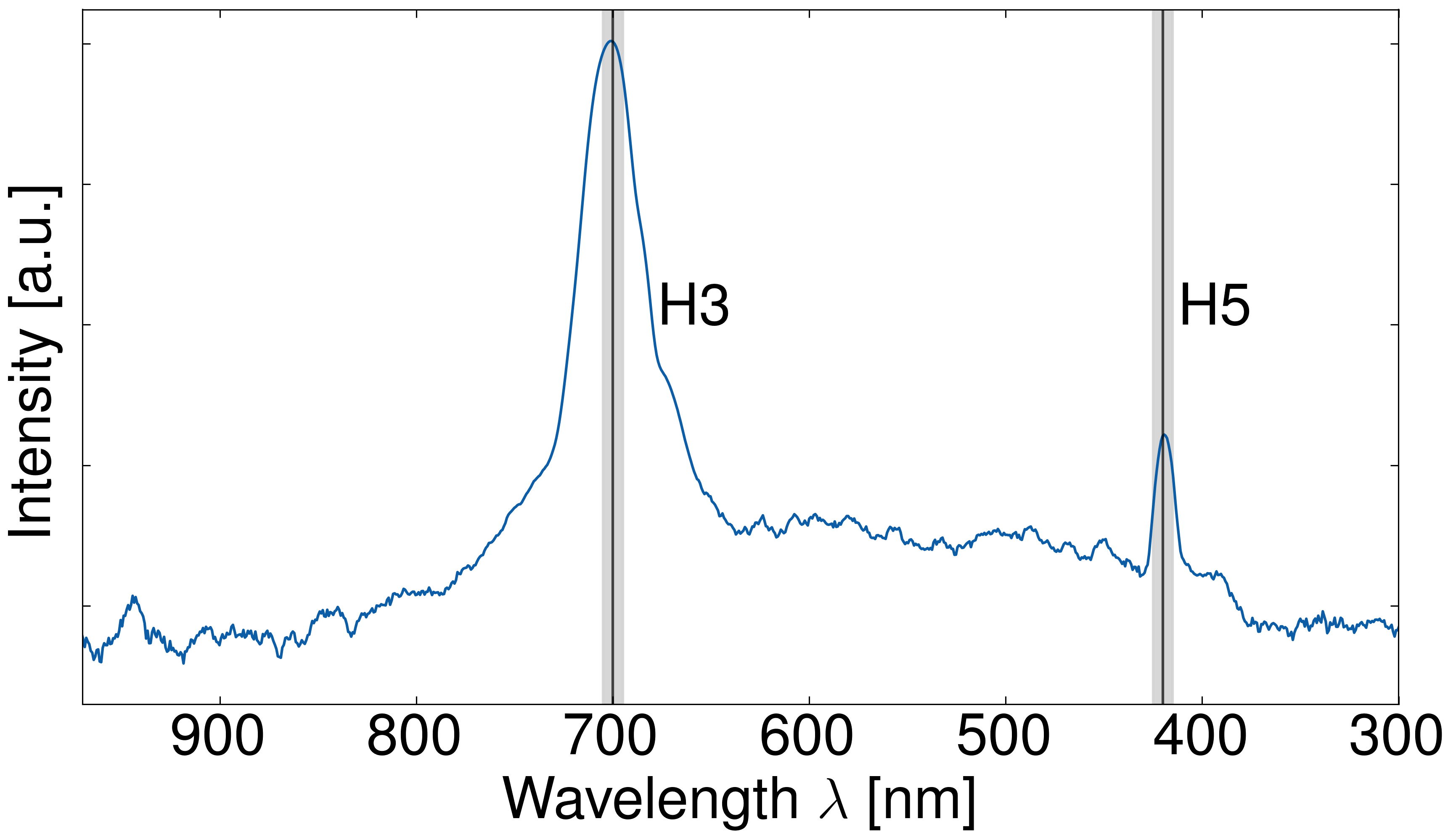}
       \end{subfigure}
              \begin{subfigure}{\textwidth}
       \subcaption{}
            \includegraphics[width=\linewidth]{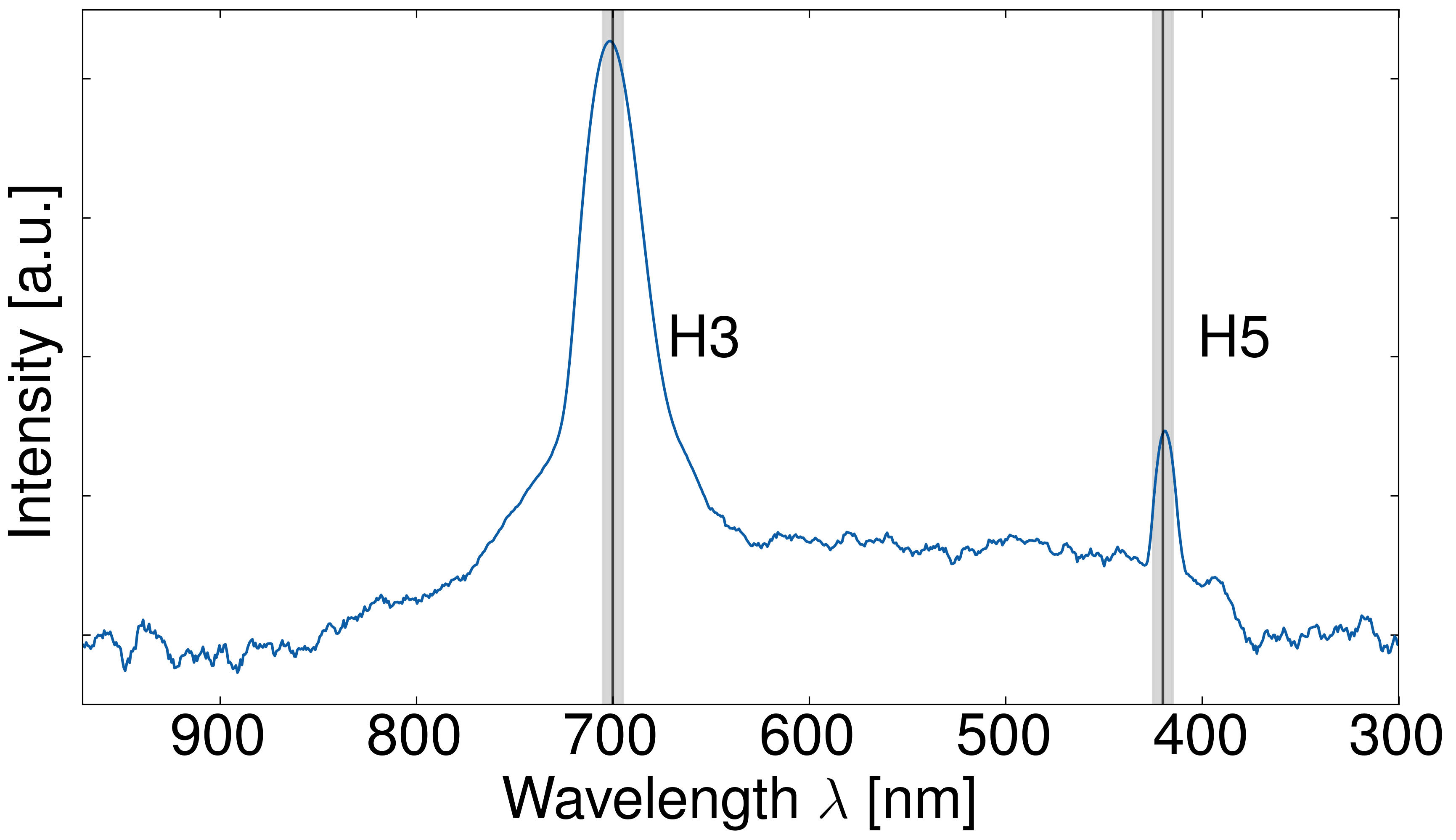}
       \end{subfigure}
       \end{minipage}%

     \centering

     \caption{\textbf{Spectra from three different semiconductors.} Measured spectra for the \textbf{a)} GaAs, \textbf{b)} ZnO and \textbf{c)} Si sample. Clearly distinguishable from the luminescence background are the intensity peak corresponding to the third and fifth high-harmonic generated in the respective sample. The gray region marks the spectral width of the employed spectral filters for the correlation measurement. The intensity axis is depicted with a logarithmic scale.}
     \label{fig:spectra}
\end{figure}

The spectra obtained from all the investigated materials are shown in Fig. \ref{fig:spectra}. The presence of the third and fifth harmonic order is clearly visible and the width of the used spectral filters is displayed for clarity. Most notably, no signal is detected at the position of the fourth harmonic, as it would be expected from a perturbative generation process. Also no signal is detected at the wavelength of the second harmonic, which is not included in the depicted spectra as the measurement was performed with a different spectrometer.

 Additionally to the power scaling law and the spectra, four parameters are of equal importance to require nonlinear ionization,
 The Keldysh parameter can indicate which generation process dominates the HHG process in a given material. 
For $\gamma_\mathrm{K} > 1$ it is assumed that below band-gap, perturbative harmonics are generated, while for $\gamma_\mathrm{K} < 1$ non-perturbative HHG is usually the main emission channel. 
In the GaAs sample, we calculate $\gamma_\mathrm{K} = 0.616 < 1$ for the highest driving laser intensity, using $m^* = 0.067$, $E_\mathrm{Gap} = 1.424\, \text{eV}$ and a laser peak intensity of $I = 0.15\, \text{TW} / \text{cm}^2$ and a measured beam waist of $5$ µm. 
For measurements in the Si and ZnO crystals, $\gamma_\mathrm{K} > 1$. Still, $\hbar \omega_{\mathrm{IR}} = 0.59\, \text{eV} \ll E_{\mathrm{Gap}}$, where $E_{\mathrm{Gap}}$ the bandgap of the material, $\omega_{\mathrm{IR}}$ the angular frequency of the driving laser beam and $\hbar$ the reduced Planck constant.
 Further, for all materials the dipole approximation is valid as $z / 2c \ll 1$ with $z \equiv  U_\mathrm{p} / \omega_{\mathrm{IR}}$, $c$ the speed of light and $U_\mathrm{p}$ the ponderomotive potential. 
Lastly, $z_\mathrm{f}  \equiv 2 U_\mathrm{p}/m_\mathrm{e} c^2 $ shows that the electron’s behavior is non-relativistic, as it evaluates to $z_\mathrm{f} \ll 1$ for all materials as it is required for the tunneling regime. Concluding, the investigated sample of GaAs shows clear non-perturbative signatures for H3 and H5. For the investigated samples ZnO and Si, the non-perturbative nature is less pronounced, but still apparent as an intensity crossover. Although the Keldysh parameter is above 1, we confirm the validity of the dipole approximation employed by our theoretical model and as well the non-relativistic behavior of the electron. Most importantly, our employed theory approximations hold for all materials.

\begin{figure}[H]
     \centering
           \includegraphics[scale=0.4]{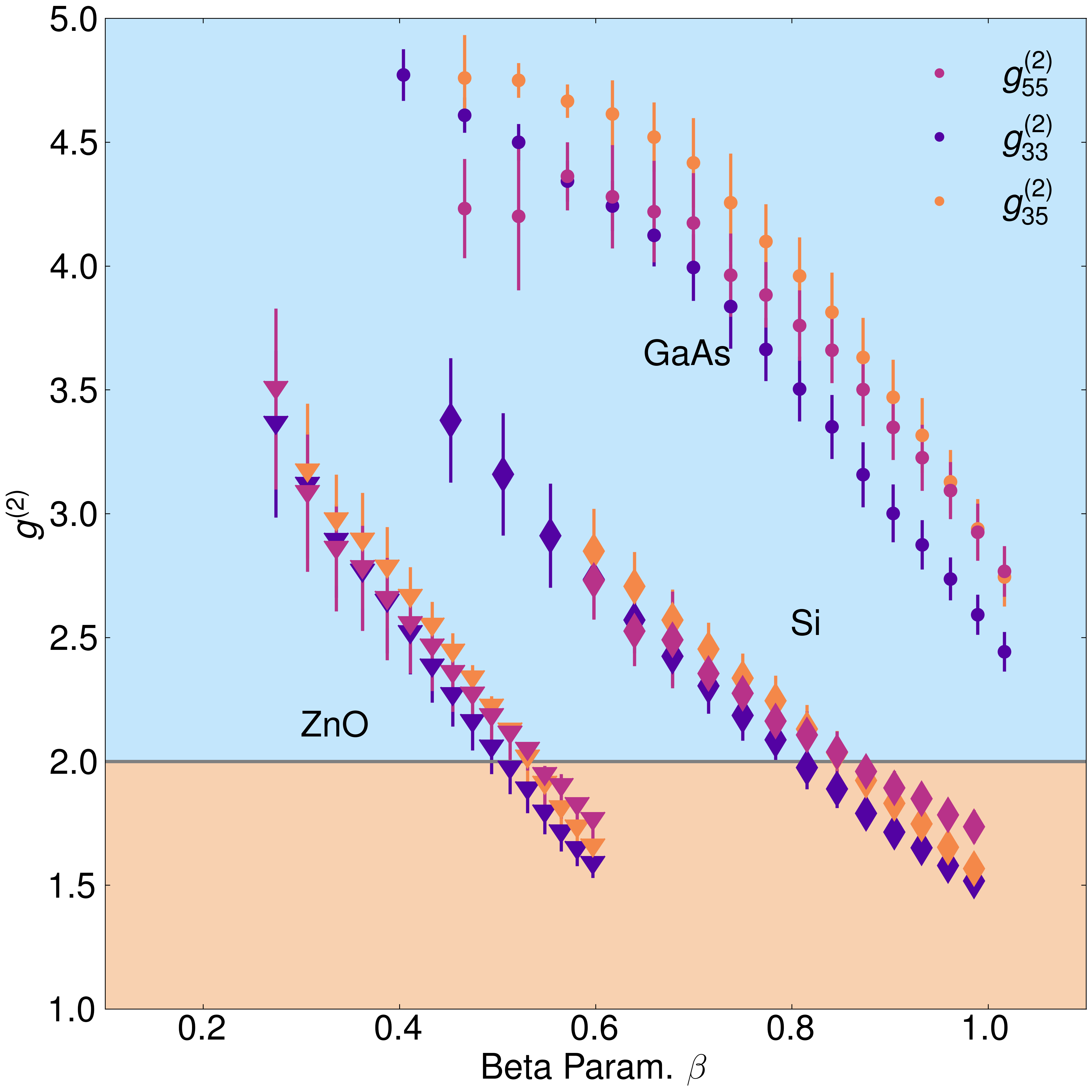}

     \caption{\textbf{Joined results from three different semiconductors.} Experimentally obtained single- and double beam second order correlation values from the third and fifth high-harmonic generated in three different crystals depicted over the Bloch parameter $\beta = \omega_B / \omega_L$, which is the ratio of the Bloch frequency of the electrons $\omega_B$ to the driving laser angular frequency $\omega_L$. Notably, the measured values for Si and ZnO are almost similar. The photon correlations obtained from the GaAs sample shows a higher magnitude of bunching. The blue colored region indicates values of super-bunching with $g^{(2)} > 2$.}
     \label{fig:beta_parameter_all}
\end{figure}

Overall, comparing the second order correlation function $g^{(2)}$ from Silicon and Zinc Oxide to the Gallium Arsenide semiconductor crystal, the slope of $g^{(2)}$ increases with decreasing Keldysh parameter. We observe similar results for ZnO and Si (Fig. \ref{fig:g2_keldysh_all}, Fig. \ref{fig:beta_parameter_all}), which also have a comparable Keldysh parameter. A possible explanation could be related to the laser-driven dependence of the non-linear material response governed by the material dependent properties, like the effective electron mass and the bandgap.

\textbf{Origin and nature of HHG squeezing.} 
For our analysis, we adopt the derivation of the quantum properties of intraband harmonics studied in the interaction picture \cite{gonoskov2022nonclassical}.
 
The observed behaviour of the system under measurement is influenced by an interaction wherein electrons respond to the oscillating vector potential of a laser field. This interaction occurs within a semiconductor material, where the electrons reside within a band structure characterized by the energy dispersion relationship, denoted as $E_c(\Vec{K})$. 
In general, recombinations of electrons are also possible source of higher harmonics above the band gap (interband contribution). However, we will mainly focus on the intraband currents. Also, in case interband harmonics might play a more important role, it will be sufficient in a first investigation to restrict our analysis on the intraband dynamics due to the following argument. In case the interband mechanism generates radiation with no or only negligible nonclassical properties, we are still capturing relevant physics. In case interband mechanism generates radiation with nonclassical properties, we are capturing only parts of the nonclassical physics, which is qualitatively satisfying.

\begin{figure}[H]
    \centering
    \includegraphics[scale=0.85]{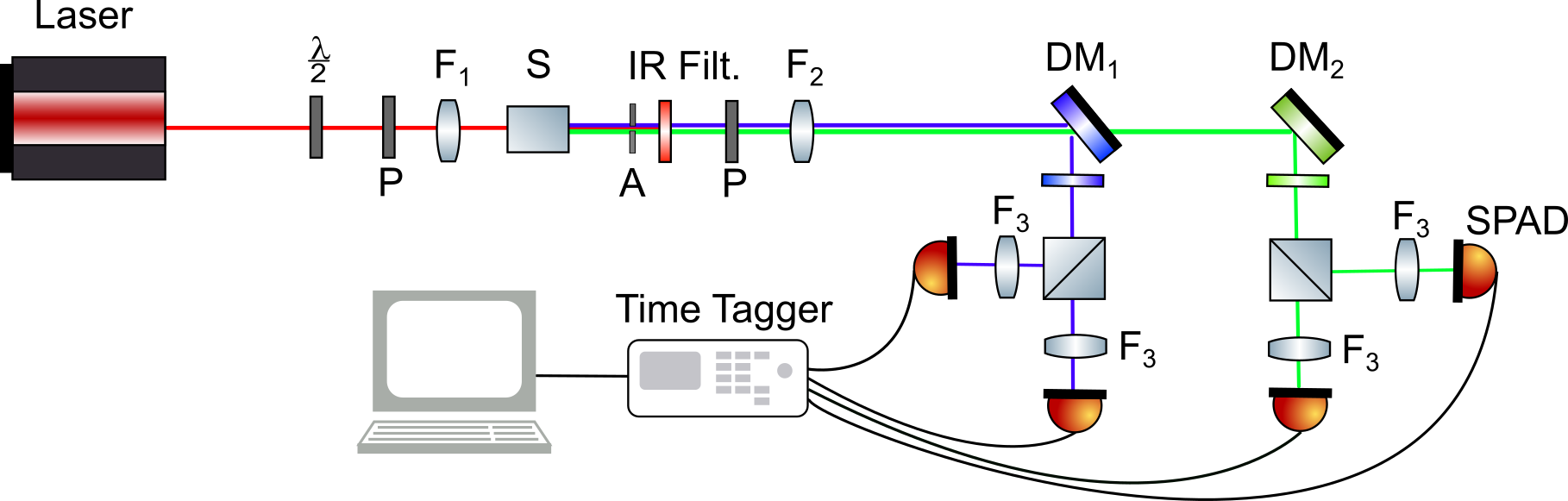}
    \caption{\textbf{Experimental configuration for intensity correlation measurements} Setup for measuring the single and double beam second order intensity correlation of HHG from a semiconductor crystal. Ultrashort pulses from the SWIR laser system are passed through a half-wave plate and a polarizer (P) and focused with a lens ($\text{F}_1$) on the semiconductor sample (S), reaching an electric field strength comparable to the material inner atomic field strength at the focus. The generated radiation is filtered spatially by an aperture (A) and selected along the main emission polarization (P) axis. The remaining infrared pump photons are filtered out. A lens ($\text{F}_2$) is employed to collimate the selected HHG radiation towards the detector arms. After, H3 and H5 are separated spatially with two dichroic mirrors ($\text{DM}_1$ and $\text{DM}_2$). Further spectral filtering is done by narrowband filters before the HBT like setup to correlate the photon arrival times. Finally, two similar lenses ($\text{F}_3$) focus the radiation on the SPAD chips. The SPADs are operated in Geiger Mode and serve as input for the Start-Stop measurement mediated by a Time-to-Digital Converter.}
    \label{fig:setup}
\end{figure}

Here, we perform additional approximations specific to our experimental conditions. We include second order terms with respect to the field operators for the harmonic modes in the interaction Hamiltonian. This is motivated by an effective expansion of the Hamiltonian in terms of the ratio of the operator norm of the field operators of the harmonic fields over the operator norm of the fundamental laser mode. 
This matches with the theoretical classification in terms of the Keldysh parameter $\gamma_\mathrm{K}$ as well as the Bloch parameter (Fig. \ref{fig:beta_parameter_all}) \cite{gonoskov2022nonclassical}. 
Furthermore, the system is reduced to a bipartite state $|\Psi\rangle_{35}$, consisting of the third and fifth harmonic selected optically (Fig. \ref{fig:power_scaling_GaAs} and Fig. \ref{fig:setup}).

The single particle Hamiltonian describing the interaction is written as
\begin{equation}
    \hat{H} = n_\mathrm{e} E_\mathrm{c}\left[\hat{\vec{p}} - \frac{e}{c} \hat{\vec{A}}\right] + \omega_3 \hat{n}_3 + \omega_5 \hat{n}_5
\end{equation}
where $e$ is the elementary charge, $c$ the speed of light, $n_e$ the number of electrons, $\hat{\vec{p}}$ the momentum of the electrons, $\hat{n}_j = \hat{a}_j^\dagger \hat{a}_j$ the photon number operator of harmonic mode $j$ with frequency $\omega_j$, $E_c$ the conduction band dispersion and $\hat{\vec{A}} =\vec{A}_{\mathrm{L}} + \hat{\vec{A}}_3 + \hat{\vec{A}}_5$ the vector potential operator of the entire radiation field, where the undepleted driving laser field is treated as a classical field.
By means of the induced momentum of the driving laser field, the electrons explore up to 30 $\%$ of the Brillouin zone. Thus, treating the canonical momentum of the Bloch electrons as dominated by the driving laser field
($\| \hat{\vec{p}} \| \ll e/c \| \hat{\vec{A}} \| $)
 and switching to the interaction picture yields the
the time-dependent interaction Hamiltonian
\begin{equation}
    \hat{H}_\mathrm{I} = n_\mathrm{e} E_\mathrm{c}\left[-\frac{e}{c} \hat{\vec{A}}(t) \right] \quad .
\end{equation}
The time dependence of the radiation field operator $\hat{\vec{A}}(t)$ is due to the fact that we are working in the interaction picture. 
We focus on the polarization projection, which allows us to treat the vector potential as a scalar quantity, and approximate the band dispersion as a cosine potential. Therefore, the time dependence of the individual modes is given as $\hat{A}_j(t) = \exp\left(i\omega_j \hat{n}_j\right) \hat{A}_j \exp\left(-i\omega_j \hat{n}_j\right)$. Substituting\\
 the potential operator with the respective field operators, the second order expansion around $-e/c A_\mathrm{L}$ gives
\begin{samepage}
\begin{align}
     \hat{H}_\mathrm{I}(t) = 
     n_\mathrm{e} E_\mathrm{g} \bigg\{ &1 - \cos\left(\frac{\pi \tilde{A}_\mathrm{L}}{K_\mathrm{c}}\right) 
    + \sin\left(\frac{\pi \tilde{A}_\mathrm{L}}{K_\mathrm{c}}\right)  
    \Big[ \tilde{c}_3 \Big(\hat{a}_3(t) + \hat{a}_3^\dagger(t)\Big) + \tilde{c}_5 \Big(\hat{a}_5(t) + \hat{a}_5^\dagger(t)\Big) \Big] \notag\\
    &+  \frac{1}{2} \cos\left(\frac{\pi \tilde{A}_\mathrm{L}}{K_\mathrm{c}}\right) 
    \Big[ \tilde{c}_3^2 \Big(\hat{a}_3(t) + \hat{a}_3^\dagger(t)\Big)^2 +  \tilde{c}_5^2 \Big(\hat{a}_5(t) + \hat{a}_5^\dagger(t)\Big)^2 \notag\\
    &+ 2\tilde{c}_3 \tilde{c}_5 \Big(\hat{a}_3(t) + \hat{a}_3^\dagger(t)\Big) \Big(\hat{a}_5(t) + \hat{a}_5^\dagger(t)\Big)
    \Big] \bigg\} ,
\end{align}
\end{samepage}
 with $\hat{a}_j^\dagger(t)$ and  $\hat{a}_j(t)$ the creation and annihilation operator respectively of high-harmonic mode $j = 3, 5$. The time dependence of the annihilation and creation operators of the individual modes is $\hat{a}_j(t) = \hat{a}_j\exp\left(-i\omega_jt\right)$, where the time independent operators are the ladder operators in the Schrödinger picture. Further, $\tilde{c}_j = \pi / K_\mathrm{c} \cdot \sqrt{\pi e^2 / \omega_j V}$ with $K_\mathrm{c}$ an inverse lattice constant, $E_\mathrm{g}$ the conduction band half-width,  $V$ the quantization volume. The classical laser potential is denoted as $\tilde{A}_\mathrm{L} = e/c A_\mathrm{L}$ with $c$ the speed of light.

The first two terms on the right-hand-side of Eq.~\eqref{eq:hamiltonian} are independent of the field operators; thus, they induce a shift of the vacuum energy. The remaining terms in the first line depend linearly on the ladder operators in such a way that a coherent displacement of the harmonic modes is induced. In a first approximation, we assume that this contribution to the Hamiltonian is negligible due to the low laser intensities. Therefore, the terms separated in the second and third line on the right-hand-side of Eq.~\eqref{eq:hamiltonian} dominate the creation of the harmonic modes. Further, we can rewrite this part in the following way:

\begin{equation}
(\hat{a}_j(t) + \hat{a}_j^\dagger(t))^2 = {\hat{a}_j(t)}^2 + {\hat{a}_j^\dagger(t)}^2 + 1 + 2 \hat{n}_j \quad ,
\label{eq{:}single_mode}
\end{equation}
where $\hat{n}_j = \hat{a}_j^\dagger \hat{a}_j$ is the photon number operator. The quadratic dependencies of the field operators in Eq.~\eqref{eq{:}single_mode} are equivalent to an interaction Hamiltonian inducing single-mode squeezing in each harmonic mode. The last contribution to the interaction Hamiltonian has the form
\begin{equation}
(\hat{a}_3(t) + \hat{a}_3^\dagger(t))(\hat{a}_5(t) + \hat{a}_5^\dagger(t)) = \hat{a}_3(t)\hat{a}_5(t) + \hat{a}_3^\dagger(t)\hat{a}_5^\dagger(t) + \hat{a}_3(t) \hat{a}_5^\dagger(t) + \hat{a}_3^\dagger(t) \hat{a}_5(t)
\label{eq{:}two_mode}
\end{equation}
that has contributions being equivalent to a two-mode squeezing Hamiltonian between the third and fifth harmonic. The last two terms in Eq. \eqref{eq{:}two_mode} further mix the two harmonic modes in a nontrivial way.
Summarizing, the theoretical treatment of HHG in an intraband process gives rise to single-beam and dual-beam squeezing of the generated spectral modes. Other effects like coherent displacement also influence the generation process, but are negligible for our experimental parameters.

\textbf{Influence of multimode detection.}
Our detection is sensitive to the effective number of squeezing modes. The reason for this is, that the time resolution of our system measuring the normalized second order correlation function is well above the coherence time of the radiation. An estimated upper limit for the coherence time is the duration of the driving laser pulse of 80 fs. Meanwhile, the timing resolution of the detection is on the order of tens of picoseconds, and therefore a time-integrated correlation function
\begin{equation}
   g^{(2)} = \frac{\int d t_1 \int d t_2 \langle \hat{a}^\dagger(t_1)\hat{a}^\dagger(t_2)\hat{a}(t_1)\hat{a}(t_2)\rangle}{\int d t_1 \langle \hat{a}^\dagger(t_1)\hat{a}(t_1)\rangle\int d t_2 \langle \hat{a}^\dagger(t_2)\hat{a}(t_2)\rangle} 
\end{equation}
is measured for the single beam correlation \cite{christ2011probing, zielnicki2018joint}. Transforming the $g^{(2)}$ function into the frequency domain by means of Fourier transformation 
\begin{equation}
         g^{(2)} = \frac{\int d \omega_1 \int d \omega_2 \langle \hat{a}^\dagger(\omega_1)\hat{a}^\dagger(\omega_2)\hat{a}(\omega_1)\hat{a}(\omega_2)\rangle}{\int d \omega_1 \langle \hat{a}^\dagger(\omega_1)\hat{a}(\omega_1)\rangle\int d \omega_2 \langle \hat{a}^\dagger(\omega_2)\hat{a}(\omega_2)\rangle} 
\end{equation}
and further transforming into a broadband basis yields
\begin{equation}
      g^{(2)} = \frac{\left\langle{:}\left(\sum_k \hat{A}^\dagger_{k} \hat{A}_{k}\right)^2{:}\right\rangle}  {\left\langle\sum_k \hat{A}^\dagger_{k} \hat{A}_{k} \right\rangle^2}  
      \label{eq:g2_broad}
\end{equation}
where the summation $k$ runs over the measured modes and $\langle {:} {:} \rangle$ denotes the normal ordering of the operators. 

From the derived interaction Hamiltonian Eq. \eqref{eq:hamiltonian}, it is apparent, that the high-harmonic radiation can exhibit single-mode and two-mode squeezing. As our measurements show a strong correlation in the double beam measurements (Fig. \ref{fig:power_scaling_ZnO}) we can treat this as the dominating effect.

Taking into account, that the broadband, multimode detection of the second order correlation Eq. \eqref{eq:g2_broad}, is sensitive to the squeezer mode distribution $\lambda_k$ and gain $B$ by employing the general Heisenberg representation of a multimode twin-beam squeezer as 
\begin{equation}
    g^{(2)} = 1 + \frac{\sum_k \sinh^4(\lambda_k B)}{ \left[ \sum_k \sinh^2(\lambda_k B)\right]^2 }
\end{equation}
 we can explain the measured scaling of $g^{(2)}_{ij}$ with the driving laser intensity in Fig. \ref{fig:power_scaling_ZnO} \cite{christ2011probing}. As the driving laser intensity increases, the gain $B$ as well as the number of occupied excited squeezing modes $k$ increases. The broadband detection then measures the convolution of the statistics of all excited modes, resulting in a Poissonian photon-number distribution yielding $g^{(2)} = 1$. 

Considering that we operate in a low-gain regime to restrict the number of events, we can connect the measured second order correlation to the effective number of squeezing modes.
\begin{equation}
    g^{(2)} = 1 + \frac{1}{K} \quad ,
    \label{eq:g2_schmidt}
\end{equation}
where $K = 1 / \lambda_k^4$ is the Schmidt number quantifying the number of effective modes measured \cite{christ2011probing}.
For low driving laser intensities, we see that $K \approx 1$, corresponding to a single twin-beam squeezer. As the driving laser intensity is increased, a greater amount of squeezing modes is occupied, resulting in $K > 1$ and a decrease in  $g^{(2)}$. Further, the change in the double beam correlation $g^{(2)}_{35}$ can be associated with an increase of the gain \cite{christ2011probing}. 
For low driving laser powers, our experimental data shows that $g^{(2)}$ is greater than 2, which exceeds the limit imposed by Eq. \eqref{eq:g2_schmidt} for the case where the effective number of modes $K = 1$.
The reason for the deviation from the employed model is, that Eq. \eqref{eq:g2_schmidt} is based on broadband, multimode detection of squeezers, specifically for a twin-beam squeezer with the same number of photons in each beam.
In our experiment, the number of photons between different harmonic modes differs significantly, deviating from the theoretical model. To address this issue, correlating two high-harmonic modes from the so-called plateau regime, with comparable intensities, would be a suitable approach. Additionally, selecting only a few spectral modes would reduce the number of contributing modes and make the scaling behaviour more apparent.

\section{Additional Data}\label{secA1}

\begin{figure}
     \centering
     \includegraphics[width=0.9\linewidth]{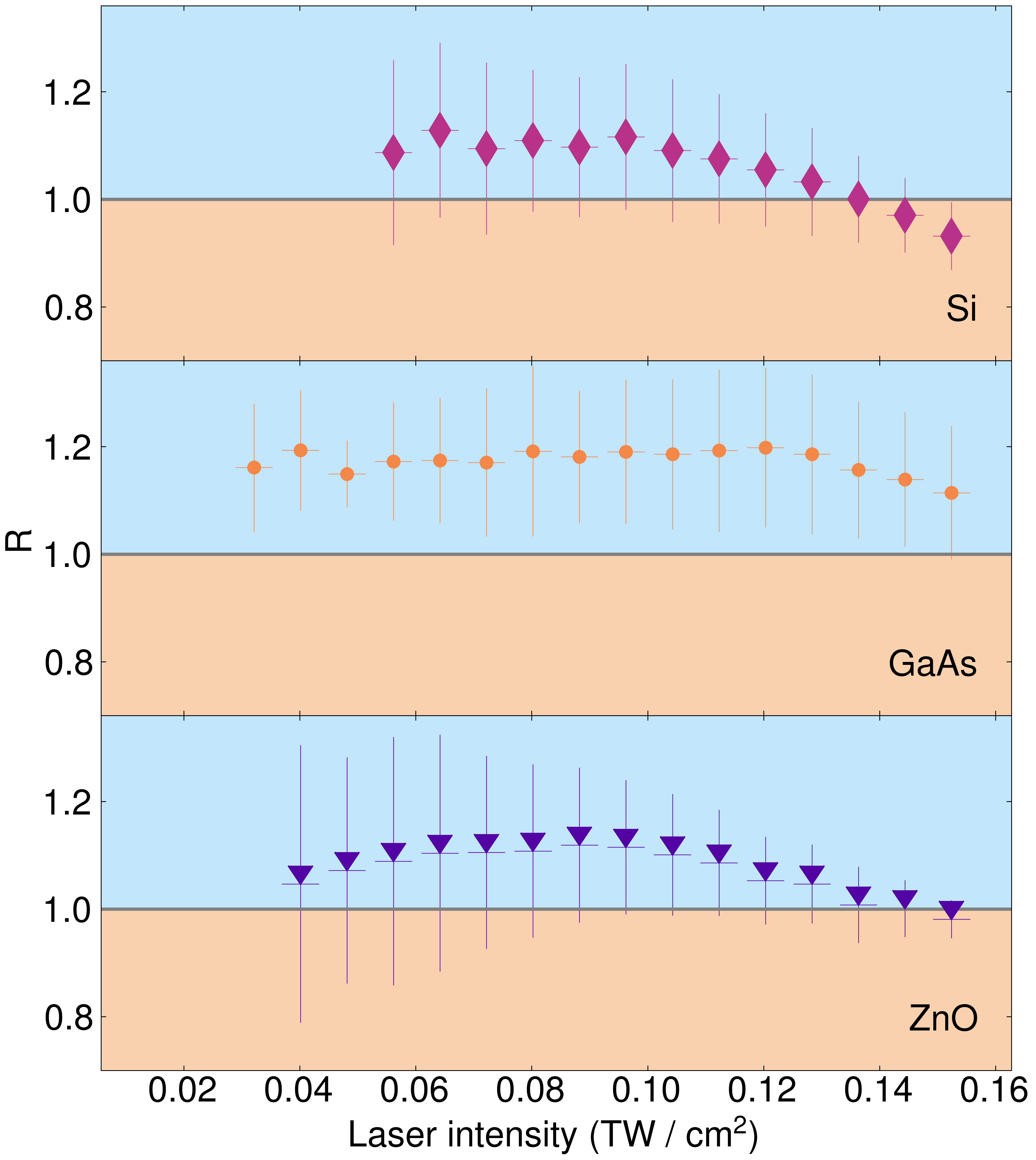}
     \caption{\textbf{Violation of multimode Cauchy-Schwarz inequality.} The graph shows the $R$ parameter, defined as in Eq. \eqref{eq:R_param}, calculated from the experimental data in Fig. \ref{fig:g2_keldysh_all} of the investigated samples. A value $R > 1$ corresponds to the violation of the classical limit given by the Cauchy-Schwarz inequality. The classical limit is depicted as the gray line. The blue colored region indicates values of non-classicality. Each point correspond to an average of a sequence of 8 measurements reproduced in the same experimental conditions. The error bars are calculated by error propagation from the original data. The graph shows that the obtained correlation values cross the classical limit of 1, indicating non-classical multimode correlations.}
     \label{fig:ncl_all}
\end{figure}

\end{appendices}



\end{document}